\documentclass[useAMS,usenatbib]{mn2e}
\usepackage{graphicx}
\usepackage{amssymb}
\usepackage[fleqn]{amsmath}
\usepackage{color}

\newcommand{\vc}[1]{\bmath{#1}} 
\newcommand{\lx}[1]{\rmn{#1}}
\newcommand{\tns}[1]{\textbfss{#1}}
\newcommand{\pdif}[2]{\frac{\partial {#1}}{\partial {#2}}}
\newcommand{\odif}[2]{\frac{\lx{d}{#1}}{\lx{d} {#2}}}
\newcommand{\odbldif}[2]{\frac{\lx{d}^2{#1}}{\lx{d} {#2}^2}}
\newcommand{\vr}[0]{\hat{\vc{e}}_r}
\newcommand{\vz}[0]{\hat{\vc{e}}_{z}}
\newcommand{\vphi}[0]{\hat{\vc{e}}_{\phi}} 

\newcommand{\li}[0]{\lx{i}}

\newcommand{\Bphi}[0]{\alf_\phi} 
\newcommand{\Bz}[0]{\alf_z}

\newcommand{\alf}[0]{B} 
\newcommand{\pert}[1]{\delta{#1}} 
\newcommand{\dur}[0]{\pert{u_r}}
\newcommand{\duphi}[0]{\pert{u_\phi}}
\newcommand{\duz}[0]{\pert{u_z}}
\newcommand{\dbr}[0]{\pert{\alf_r}}
\newcommand{\dbphi}[0]{\pert{\alf_\phi}}
\newcommand{\dbz}[0]{\pert{\alf_z}}
\newcommand{\gam}[0]{Z}
\newcommand{\dgam}[0]{\pert{\gam}}
\newcommand{\nub}[0]{\nu_{\lx{B}}}
\newcommand{\Balbus}[0]{S_B}
\newcommand{\Reynolds}[0]{\lx{Re}}
\newcommand{\Rem}[0]{\Reynolds^{\!-1}}
\newcommand{\nn}[0]{\lx{n}}
\newcommand{\vn}[0]{v_\lx{n}}
\newcommand{\kz}[0]{k}
\newcommand{\rzero}[0]{r_0}
\newcommand{\sound}[0]{c_s}

\newcommand{\gnm}[0]{\gamma_{\lx{m}}}
\newcommand{\gmn}[1]{\gamma_{\lx{m},{#1}}}
\newcommand{\gn}[0]{\gamma}

\newcommand{\oa}[0]{\sigma}
\newcommand{\oam}[0]{\sigma_{\lx{m}}}

\newcommand{\km}[0]{k_{\lx{m}}}

\newcommand{\wm}[1]{\zeta_{\lx{m}}}
\newcommand{\wmn}[1]{\zeta_{\lx{m},#1}}
\newcommand{\sntp}[1]{\sin^{#1}\theta}
\newcommand{\cstp}[1]{\cos^{#1}\theta}
\newcommand{\cotp}[1]{\cot^{#1}\theta}
\newcommand{\vbb}{\vc{b}}
\newcommand{\vu}{\vc{u}}			 
\newcommand{\vB}{\vc{\alf}}			 
\newcommand{\nuii}{\nu_{ii}}
\newcommand{\vthi}{v_{\lx{th}i}}
\newcommand{\mfp}{\lambda_{\lx{mfp}}} 

\newcommand{\ff}{a}			 
\newcommand{\rr}{b}
\newcommand{\jj}{d}
\newcommand{\ee}{E}			 
\newcommand{\cc}{C}
\newcommand{\dd}{D}

\newcommand{\eps}{\epsilon}			 
\newcommand{\order}{\mathcal{O}}			 
\newcommand{\om}{\Omega}
\newcommand{\cyc}{\omega_i}

\newcommand{\lt}{\left}		
\newcommand{\rt}{\right}			 
\newcommand{\eqn}[1]{equation (\ref{#1})}
\newcommand{\alspace}[1]{ \hfill {#1} \; }
\newcommand{\eqna}[2]{equations (\ref{#1}) and (\ref{#2})}
\newcommand{\eqnt}[2]{equations (\ref{#1}) to (\ref{#2})}
\newcommand{\bea}{\begin{eqnarray}}
\newcommand{\eea}{\end{eqnarray}}
\newcommand{\beq}{\begin{equation}}
\newcommand{\eeq}{\end{equation}}

\newcommand{\mnras}[0]{MNRAS}\newcommand{\apj}[0]{ApJ}\newcommand{\apjs}[0]{ApJ}\newcommand{\apjl}[0]{ApJ
  Letters}\newcommand{\araa}[0]{ARA\&A}\newcommand{\aap}[0]{A\&A}

\title[ Galactic MRI with Braginskii viscosity ]{Global MRI with Braginskii viscosity in a galactic profile}
\author[M. S. Rosin, A. J. Mestel]{M. S. Rosin$^{1,2}$\thanks{E-mail:
    msr35@math.ucla.edu} and  A. J. Mestel $^{3}$\\
  $^{1}$ UCLA Department of Math, 520 Portola Plaza, Los Angeles, CA 90095, U.S.A.\\
  $^{2}$ DAMTP, Centre for Mathematical Sciences, University of
  Cambridge, Wilberforce Road,
  Cambridge,  CB3 0WA, U.K.\\
  $^{3}$ Department of Mathematics, Imperial College, South Kensington
Campus, London, SW7 2AZ, U.K.}
\begin{document}

\date{Submitted to MNRAS April 2012}

\pagerange{\pageref{firstpage}--\pageref{lastpage}} \pubyear{2011}

\maketitle

\label{firstpage}


\begin{abstract}
We present a global-in-radius linear analysis of the  axisymmetric magnetorotational instability (MRI) in 
 a collisional magnetized plasma with 
Braginskii viscosity.  For a galactic angular velocity profile $\Omega$ we  obtain analytic solutions for
 three magnetic field orientations: purely azimuthal, purely vertical and slightly pitched (almost azimuthal). In the first two cases the Braginskii viscosity damps otherwise neutrally stable modes, and reduces the growth rate of the MRI respectively.
  In the final case the Braginskii viscosity makes the MRI up to $2\sqrt{2}$ times faster  than its inviscid counterpart, even for \emph{asymptotically small} pitch angles. We investigate the transition between the Lorentz-force-dominated and the Braginskii viscosity-dominated regimes in terms of
a parameter $\sim \Omega \nub/B^2$ where $\nub$ is the viscous coefficient and $B$ the Alfv\'en speed. In the limit where the parameter
is  small and large respectively we recover the inviscid MRI and the magnetoviscous instability (MVI).  We obtain asymptotic expressions for the
approach to these limits, and find the Braginskii viscosity can magnify the effects of azimuthal hoop tension (the growth rate becomes complex) by over an order of magnitude. We discuss the relevance of our results to the local approximation, galaxies and other magnetized astrophysical plasmas.  Our results should prove useful for benchmarking codes in global geometries.  

\end{abstract}




\section{Introduction} \label{sec:intro}

In the formation of compact objects (stars, planets and black holes)
from accretion discs, turbulence driven by the MRI, and possibly the MVI,
offers a promising mechanism for the necessary angular momentum
transport \citep{Velikhov_59, Chand_60, Balbus_91, Balbus_04}. It has also been suggested
that the observed velocity fluctuations $\sim 6 \,\lx{km\,s}^{-1}$ in
 parts of the  interstellar medium (ISM) with low star formation rates may,
in part, arise from this process
\citep{Sellwood_99, Tamburro_09}.  The evidence for this comes
primarily from numerical simulations and a wide range of studies agree
that weak magnetic fields and outwardly decreasing angular velocity
profiles are an unstable combination \citep{Balbus_98_rev, Balbus_03_rev}

The most illuminating explanation for this comes from a shearing sheet
analysis in which the mean azimuthal flow of the differentially
rotating disc is locally approximated by a constant angular velocity
rotation plus a linear velocity shear \citep{Goldreich_65,
  Umurhan_04}. In the simplest possible setup, incompressible,
isothermal, dissipationless, axisymmetric linear perturbations to a
magnetic field with a weak vertical component, i.e. parallel to the rotation axis of
the disc, are unstable when the angular velocity decreases
away from the disc's centre.  Azimuthal velocity perturbations to
fluid elements at different heights, tethered to each other by the
magnetic field, increase (decrease) their angular momentum. This
causes them to move to larger (smaller) radii as dictated by the
gravitational field which sets the mean flow. In the frame rotating at
constant angular velocity, this motion deforms the tethering magnetic
field, provided it is not too strong, and this induces a prograde
(retrograde) Lorentz force on the outer (inner) element thus
destabilising the system as it moves to yet larger (smaller)
radii. This mechanism is at the heart of the MRI.

However, although this model and description captures much of the
essential physics, to fully understand the MRI, or at least the
framework within which the shearing sheet should exist, a more nuanced
approach is needed. In part, this is because the shearing sheet is
formulated in a Cartesian coordinate system where curvature terms that
arise naturally from the cylindrical geometry of the accretion disc
are neglected. Indeed, in the local approximation, the
  over-stabilising effects of hoop tension (a curvature effect)
  associated with the azimuthal magnetic field, are totally ignored.
Furthermore, the model predicts that the fastest growing, and
therefore most physically relevant, linear MRI modes have a homogeneous
radial structure on the scale of the shearing sheet in which the local
approximation is made ( \cite{guan09} has shown the MRI is well localized in the non-linear regime).
 This means that the global disc structure,
including boundary conditions, not captured by the local approximation
may have a significant effect on these large scale modes in a way that
cannot be determined locally. Other limitations exist too
\citep{Knobloch_92, Regev_08}.

The extent to which these limitations matter should, and under a
variety of assumptions have, been investigated by global analyses that
take into account the full radial structure of the disc and its
boundaries \citep{Dubrulle_93, Curry_94, Curry_95, Ogilvie_95,
  Ogilvie_98}, and specifically for the 
  galaxy by \cite{Kitchatinov_04}. The conclusions of these
investigations largely confirms the local picture of a large radial
scale instability driven by the differential rotation of the
disc. This suggests that whilst a local MRI analysis is generally
correct, its regime of validity must be checked globally.  It is the
purpose of this work to do just that for the MRI operating in a
collisional, \emph{magnetized} (the ion cyclotron frequency $\cyc
\gg$ ion-ion collision frequency $ \nu_{ii}$) plasma (like the ISM). 

In such a plasma \cite{Brag_65} has shown that to lowest order in
$\cyc/ \nu_{ii}$, the deviatoric stress tensor is diagonal and
anisotropic. This leads to different parallel and perpendicular
viscosities or, more fundamentally, pressures (and thermal
conductivities\footnote{In the presence of temperature gradients,
  anisotropic thermal conduction can lead to the
  magnetothermal \citep{Balbus_00} and heat flux buoyancy
  instabilities \citep{Quataert_08}.}) with
respect to the local direction of the magnetic field.  Of the
important physical consequences of this, it will be the
effect of the Braginskii viscosity in the presence of a galactic shear
flow that will concern us here.

This is not a new topic and in recent years the study of magnetized 
accretion discs has attracted attention in both the collisionless 
\citep{Quataert_02_MRI, Sharma_03_MRI, Sharma_06, Sharma_07} and collisional
regimes \citep{Balbus_04, Islam_06, Ferraro_08,Devlan_10}, and a well developed code to simulate them now exists
  \citep{Parrish_07, Stone2008athena}. However,
a number of fundamental questions remain unanswered. Primarily, what is the non-linear fate of the MRI in a collisional magnetized plasma?
Does it transport angular momentum and if so, is the transporting stress primarily 
viscous, Maxwell or Reynolds\footnote{Simulations in the collisionless regime by \cite{Sharma_06}
shows that there is angular momentum transport and the anisotropic
pressure constitutes a significant portion of the total 
stress ($\sim$ Maxwell and $\gg$ Reynolds).}?
On what scales do the most unstable modes emerge and how does this 
vary with the system parameters? In what regime 
will a local analysis become untenable and global effects (either radial or vertical) become
important \citep{Gammie_94}?
 What effect does the presence, or absence, of a net vertical field have given Cowling's anti-dynamo theorem 
and the dissipative properties of the Braginskii viscosity \citep{moffatt1978field, 
 Lyutikov_07}? Could viscous heating from the Braginskii viscosity lead to 
secondary magnetized or unmagnetized thermal instabilities \citep{Balbus_01,Quataert_08, Kunz_10b}? 
Do channel solutions, or something approaching them, 
emerge \citep{goodman1994parasitic}? If they do, the associated field growth
will generate pressure anisotropies that could feed new parasitic instabilities such as the mirror.
What would their effects be at this stage and in the inevitable turbulence where the mirror and
firehose instabilities will both arise \citep{Schek_05}?

Addressing these questions will require a two-pronged approach involving both numerical
and analytic studies. It may transpire that much of the existing work on the unmagnetized
and collisionless MRI is directly applicable, but this needs determining. As such we conduct
a global  linear stability analysis for three separate background magnetic field orientations: purely
azimuthal, purely vertical and pitched (magnetic field lines follow
helical paths on cylinders of constant radius). We embed these in a galactic rotation 
profile. In agreement with
earlier, local studies, we find that when the field has both a
vertical and an azimuthal component, a linear instability with a real
part up to $2 \sqrt{2}$ times faster than the MRI emerges
\citep{Balbus_04, Islam_06}. In contrast to local studies we also find
that it has a travelling wave component, and its growth rate depends on the viscous coefficient. 

In the presence of a vertical field, we also recover the 
standard inviscid MRI and show that upon introducing the 
Braginskii viscosity, its growth rate is always reduced. A similar
effect is found for a purely azimuthal field where we find 
that the Braginskii
viscosity damps modes that are, inviscidly, neutrally stable.

The layout of this paper is as follows.  In
Section \ref{sec:Equations}, we introduce and perturb a series of
equilibrium solutions to the Braginksii-MHD equations and this forms the basis of our global stability
analysis. Relegating the manipulation of the perturbed equations to Appendix \ref{sec:linear}, we obtain a single ODE governing
the perturbed modes. We proceed by solving this for azimuthal, vertical and pitched field orientations in 
Sections \ref{sec:azimuthal}, \ref{sec:vertical} and \ref{sec:pitched}  respectively. For
each case we contrast the
behaviour with and without Braginskii viscosity.  In Section \ref{sec:discussion} we
discuss the physical mechanism of the instability, where it may occur astrophysically, and the
relation of our results to the local approximation. Finally, we conclude in Section \ref{sec:Conc} with some thoughts on open questions 
relating to
magnetized astrophysical plasmas. 


\section[]{Global Stability Analysis}\label{sec:Equations}


\subsection[]{Governing equations}\label{sec:governing_equations}

The simplest set of equations that capture the physics of the collisional magnetized
MRI are those those of isothermal ideal MHD with the Braginskii viscosity \citep{lifshitz1984electrodynamics}.  Explicitly these equations are the momentum
equation, including the Braginskii stress tensor; the induction
equation that describes the evolution of the magnetic field; the
incompressibility condition (because the perturbations are linear, the Mach number can 
always be made small enough to ensure this); and the
stress tensor itself:
\begin{align}
  \pdif{\vc{u}}{t} + \vc{u}\cdot \nabla \vc{u} &= -\nabla\Pi + \vB\cdot\nabla\vB - \nabla \Phi_{\lx{D}}- \nabla \cdot \tns{T},\label{eq:mom2}\\
  \pdif{\vB}{t} + \vc{u}\cdot\nabla \vB &= \vB\cdot\nabla \vc{u}, \label{eq:ind}\\
  \nabla \cdot \vc{u} &= 0\frac{}{},\label{eq:incomp}\\
  \tns{T} &=\nub  \left(\tns{I} - 3\vc{b b}\right)  \vbb\vbb:\nabla \vu\label{eq:stress},
\end{align}
where the constant mass density has been
scaled out of the problem, $\vc{u}$ is the velocity field, $\vB$ is the mass-density scaled magnetic field, i.e. the  Alfv\'en velocity,
$\vbb=\vB/\alf$ is the direction of the magnetic field,
$\Phi_{\lx{D}}$ is the gravitational potential, $\Pi = p + B^2/2$ is
the total (gas plus magnetic) pressure, `:'  is the full inner product,  $\tns{I}$ the identity, and $\tns{T}$ is the full
Braginskii stress tensor whose form we now explain. 

In a magnetized plasma the total pressure tensor $\tns{P}= p \tns{I} + \tns{T}$ (whose divergence appears in the momentum equation) is given by
\beq
\tns{P} = p \tns{I} + \frac{1}{3}\lt( \tns{I} - 3 \vbb \vbb\rt) \lt( p_\perp - p_\parallel \rt), \nonumber
\eeq  
where $p_\perp, p_\parallel$ are the perpendicular and parallel scalar pressures with respect to the magnetic field direction. When the plasma is also collisional, the pressure anisotropy $p_\perp - p_\parallel$ can be related to the rate-of-strain by a Chapman-Enskog style perturbation theory \citep{Brag_65, Chap}.  

Microphysically, the anisotropy arises from
the conservation of the magnetic moment (first adiabatic invariant) of
a gyrating particle in a magnetic field. However,  collisions break this
conservation and relax the anisotropy by pitch-angle scattering
particles in velocity space (this dissipative process will turn out to
be important in isolated field configurations). The competition between these
two processes is governed by 
\beq
\frac{d}{dt} (p_\perp - p_\parallel ) \simeq 3 p \frac{d \ln B}{dt} -  \nuii  (p_\perp - p_\parallel ), \nonumber
\eeq
where we have used the BGK operator to approximate the full collision operator. 

 In the presence 
of time-varying magnetic fields, the pressure anisotropy tends to a steady state that tracks the fields' rate of change. It follows 
\beq
 p_\perp - p_\parallel = 3 \nub \frac{d \ln B}{dt} =  3 \nub  \vbb\vbb:\nabla \vu,\nonumber
\eeq
where $\nub \sim p /\nuii$   is the coefficient of the Braginskii viscosity, and we have used 
\eqn{eq:ind} in the final equality. Equation (\ref{eq:stress}) follows directly. 

\subsection[]{Equilibrium solutions}\label{sec:equilibrium}

Working in cylindrical polar coordinates $(r,\phi, z)$, we introduce equilibrium solutions to equations
(\ref{eq:mom2})-(\ref{eq:stress}) that describe a differentially
rotating global shear flow constrained by gravity and threaded by a
magnetic field that lies on cylinders of constant radius.  Our equilibrium solutions
are uniform in $z$, and  the plasma motion is restricted at an inner boundary of 
finite radius $\rzero$. (The validity of these assumptions with be discussed in Section 
\ref{sec:relevance}.) 

We allow gravity $\nabla \Phi_{\lx{D}}$ 
to dictate the rotation profile of the equilibrium flow $\om =
\om_0 (r/\rzero)^{-q}$ where $\om_0$ is the rotation frequency
at the inner boundary and $q$ is a dimensionless measure of the
shear. Of interest to us here is the case of $q=1$ that is both
analytically treatable and physically corresponds to a galactic disc
where, unlike the Keplerian case of $q = 3/2$, the gravitational
potential of the \emph{dark matter halo} sets the rotation profile
\citep{Rubin_70, Sofue_01}. In modelling this we set
the dark matter mass distribution (whose sole purpose is to ultimately
set the rotation profile) to a \cite{Mestel_63} profile so, via
Poisson's equation,  $\nabla \Phi_{\lx{D}} \propto 1/r$. In this case, the flow $\vc{u} = \om(r)
r \vphi = \om_0 \vphi$ does not vary with radius. 

We decompose the magnetic field into vertical and azimuthal components, (so
as to construct a time independent equilibrium, we neglect radial
magnetic fields\footnote{Although magnetic field configurations vary
  from galaxy to galaxy, they are commonly found tracing the spiral
  arms and therefore, in the plane of the disc, predominantly
  azimuthal.  \cite{Beck_96} gives values for the mean in-plane field
  $\alf_r/\Bphi \sim 0.25$ thereby justifying, to some degree, our
  neglect of the radial field.}). That is, $\vB =\Bz \vz + \Bphi \vphi
= \alf (\sntp{} \vz + \cstp{} \vphi)$ where $\theta =
\arctan(\Bz/\Bphi)$ is the pitch angle of the magnetic field. 
We
do not specify  $\theta$ yet, however, for mathematical simplicity, we demand that both
$\Bphi$ and $\Bz$ are independent of radius so $\theta$ remains
constant. Whilst this implies a vertical current $\propto 1/r$ is
associated with $\Bphi$, a simple super-galactic magnetic field can
account for $\Bz$. In all, our equilibrium fields take the form
\beq \label{eq:equilsoln}
\vu = \Omega_0 \vphi, \qquad \vB = \Bz \vz + \Bphi \vphi,
\eeq
and are constant in space  and time. As is physically relevant, we  restrict $B/\Omega < 1$.

It is a mathematically convenient feature of our equilibrium solutions
that there is no evolution of the magnetic field strength. From 
\eqna{eq:stress}{eq:equilsoln}, $\tns{T}$ is absent from  
 the unperturbed state and the system will be stable to pressure
anisotropy driven microscale instabilities, e.g. firehose and mirror
\citep{Schek_05}.  In contrast to the case where Laplacian viscosity
(or indeed resistivity) is present, we can construct an ideal MHD
solution independent of any radial flows \citep{Kersale_04}.


\subsection[]{Perturbed equations}\label{sec:analysis}

\begin{figure}
  \includegraphics[width=84mm]{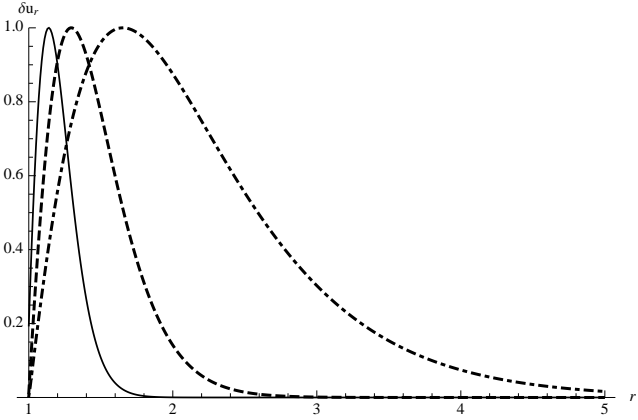}
  \caption{Radial mode structure of the
    fastest growing $\nn = 0$ branch MRI modes with $\theta = \pi/2, z= 0,
    \alf= 2.0 \cdot 10^{-2}$ and $\Balbus = 10\, \lx{(solid)},  10^2 \,\lx{(dashes)},
     10^3 \, \lx{(dots/dashes)}$. The (purely real) growth
    rates are $\gnm= 0.39, 0.31, 0.19$ and associated wavenumber  $\km = 36.0, 30.0, 20.6$.}
    \label{Fig:Mode}
\end{figure}

To determine the stability of this system we linearise equations
(\ref{eq:mom2})--(\ref{eq:stress}) about (\ref{eq:equilsoln}) with 
axisymmetric velocity perturbations 
$\delta\vc{u} = \delta\vc{u}(r) \exp[\li\kz z + \gamma t]$ and
similarly for the magnetic and pressure fields. Here $\kz$ is the
wavenumber in the z direction and $\gamma$ the growth rate. We retain
curvilinear terms from the cylindrical geometry but neglect
self-gravity. We non-dimensionalise with respect to time-scales $\om_0$
and length-scales $\rzero$.

As detailed in Appendix \ref{sec:linear}, the linearised equations combine into a single \emph{complex} second order ordinary
differential equation for $\dur$, the Modified Bessel equation
\begin{equation}\label{eq:ODE_0}
  \odbldif{\dur}{r} +  \frac{1}{r}\odif{\dur}{r} - \lt(p^2 - \frac{v^2}{r^2}\rt) \dur = 0, 
\end{equation}
where
\begin{align}
  p^2 &\equiv \frac{\kz^2}{\ee_1} (\oa^2 + \gamma^2)(\oa^2 + \gamma^2 +  \Balbus \gamma \oa^2  \cstp{2}),\label{eq:p_0}\\
  v^2 &\equiv -\frac{1}{\ee_1}(\gamma^4 + \ff_3 \gamma^3 + \ff_2 \gamma^2 +
  \ff_1 \gamma + \ff_0 ),\label{eq:v_0}
\end{align}
and
\begin{align}
  \ee_1 &= (\oa^2 + \gamma^2)(\oa^2 + \gamma^2+  \Balbus
  \gamma\oa^2),\nonumber\\
  \ff_3 &=  \Balbus \oa^2 [\cstp{2}(\cotp{2} - 1)+ \sntp{2}],\nonumber\\
  \ff_2 &= 2[\oa^2(1 + \cotp{2}) + k^2] + 2\li \Balbus \oa^2\sin 2\theta 
  \kz(1 - \cotp{2}),\nonumber\\
  \ff_1 &= \Balbus \oa^4 \lt[\cstp{2} \lt(3 - 2\kz^2\oa^{-2}\rt) + 1 -\cotp{2}\rt]- 8 \li \oa^2 \kz\cotp{},\nonumber\\ 
  \ff_0 &= \oa^2[\oa^2(1 - 2 \cotp{2}) - 2 \kz^2].\nonumber
\end{align}
Here $\oa \equiv  k B \sin \theta$ is the vertical Alfv\'en frequency and $\Balbus = 3 \nub/B^2$ (  $= 3 \nub \Omega_0/ (B^{2} n m_i)$ in dimensional form) $\simeq \beta/\nuii$ is
a measure of the relative sizes 
of the anisotropic pressure force and Lorenz force\footnote{Using a Landau fluid closure and including the effect of collisions \cite{Quataert_02_MRI, Sharma_03_MRI}
were the first to show the dependance of the MRI on $\Omega, \nuii$ and $\beta$. Because 
the closure for the pressure anisotropy differs between the collisionless and Braginskii regimes, the 
transition between the pressure anisotropy driven MRI, and the MHD MRI scales, in dimensional form, 
as $(\nuii/\Omega) \beta^{1/2}$ there, and $(\nuii/\Omega) \beta$ here. }.  Here $n$ is the number density, $m_i$ the ion mass  and $\beta$ the plasma-beta. 

To obtain solutions to \eqn{eq:ODE_0}, we choose our boundary
conditions to be an impenetrable wall at $\rzero$ so $\dur(\rzero) =
0$ and the requirement $r^{\frac{1}{2}} \dur$ decays at infinity (see
\cite{Furukawa_07} for an inviscid treatment that includes the shear
singularity at the origin).

In this case, the eigenfunctions of \eqn{eq:ODE_0} are modified Bessel functions of
the second kind $\lx{K}_{\li v}(p r)$ with argument $pr$ and order $v$
\citep{Watson, Abram}.  The spectrum of solutions is discrete and we index
$v$ with $\nn$, $\vn$. In the
special case when $p$ (or $\gamma$ from equation (\ref{eq:p_0})) is real
(in general it is complex) the problem is Sturm-Liouville and $\vn$ is
an infinite ordered set of eigenvalues $v_0<v_1<v_2\ldots <v_\infty$.
To determine $\vn$ and therefore $\gn$, it is necessary to solve
\begin{equation}\label{eq:bessel_root}
 \lx{K}_{\li \vn}(p) = 0.
\end{equation}
From solutions to this and \eqna{eq:p_0}{eq:v_0}, the full set of
flow, magnetic and pressure fields can be constructed,
Appendix \ref{sec:linear}. In Fig. \ref{Fig:Mode} we show the radial structure
of $\dur$ when  $\theta = \pi/2$.

In general, determining $\gn$ must be done numerically. However,
 for the most physically relevant magnetic field
configurations, the problem becomes, in part, analytically
tractable. These cases, a purely azimuthal field $\theta = 0$ (in the
galactic plane), a purely vertical field $\theta = \pi/2$ (a
super-galactic field) and a slightly pitched field $\theta \ll 1$
(very slightly out of the galactic plane), exhibit categorically
different behaviour such that $\theta \rightarrow 0, \pi/2$ are
singular limits.

 In the first case, linear perturbations are
damped; in the second, the system is unstable to the MRI but the
Braginskii viscosity reduces the maximum growth rate $\gnm$ below
the Oort-A value maximum $|\lx{d}\ln\om/\lx{d}\ln r/2| = 1/2$,
\citep{Balbus_92_oort}; in
the third, the system is unstable with
 $\gnm \to \sqrt{|\lx{d}\ln\om^2/\lx{d}\ln
r|}= \sqrt{2}$, even for \emph{asymptotically small} $\theta$ \citep{Balbus_04}.
We
present the details of these calculations now.


\section{Azimuthal Field, $\theta = 0$}\label{sec:azimuthal}

The stability of inviscid axisymmetric perturbations to a purely
azimuthal magnetic field in the presence of a shear flow are well
known. When $q <2$ and the magnetic field $\vB = \alf r^{-d} \vphi$ has $d>-1$ the system is always
stable (as ours is). When only one criterion is met, depending on the
form of the fields, the system may still remain stable by the modified
Rayleigh criterion \citep{Rayleigh_16, Michael_54,
  Chand_61}. (It is worth noting however that global
non-axisymmetric perturbations are unstable \citep{Ogilvie_95}.)

\begin{figure*}
  \includegraphics[width=84mm]{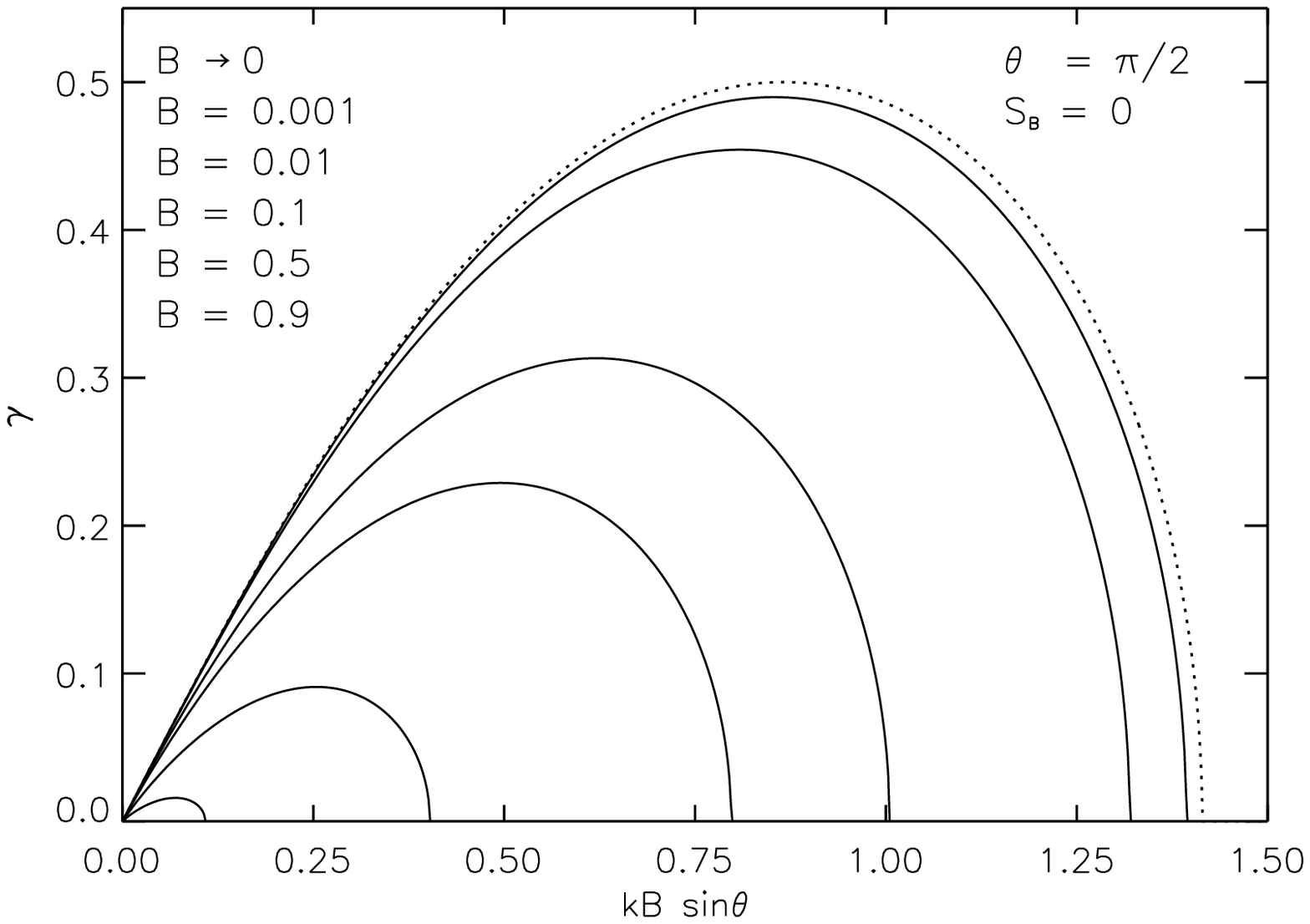}
  \includegraphics[width=84mm]{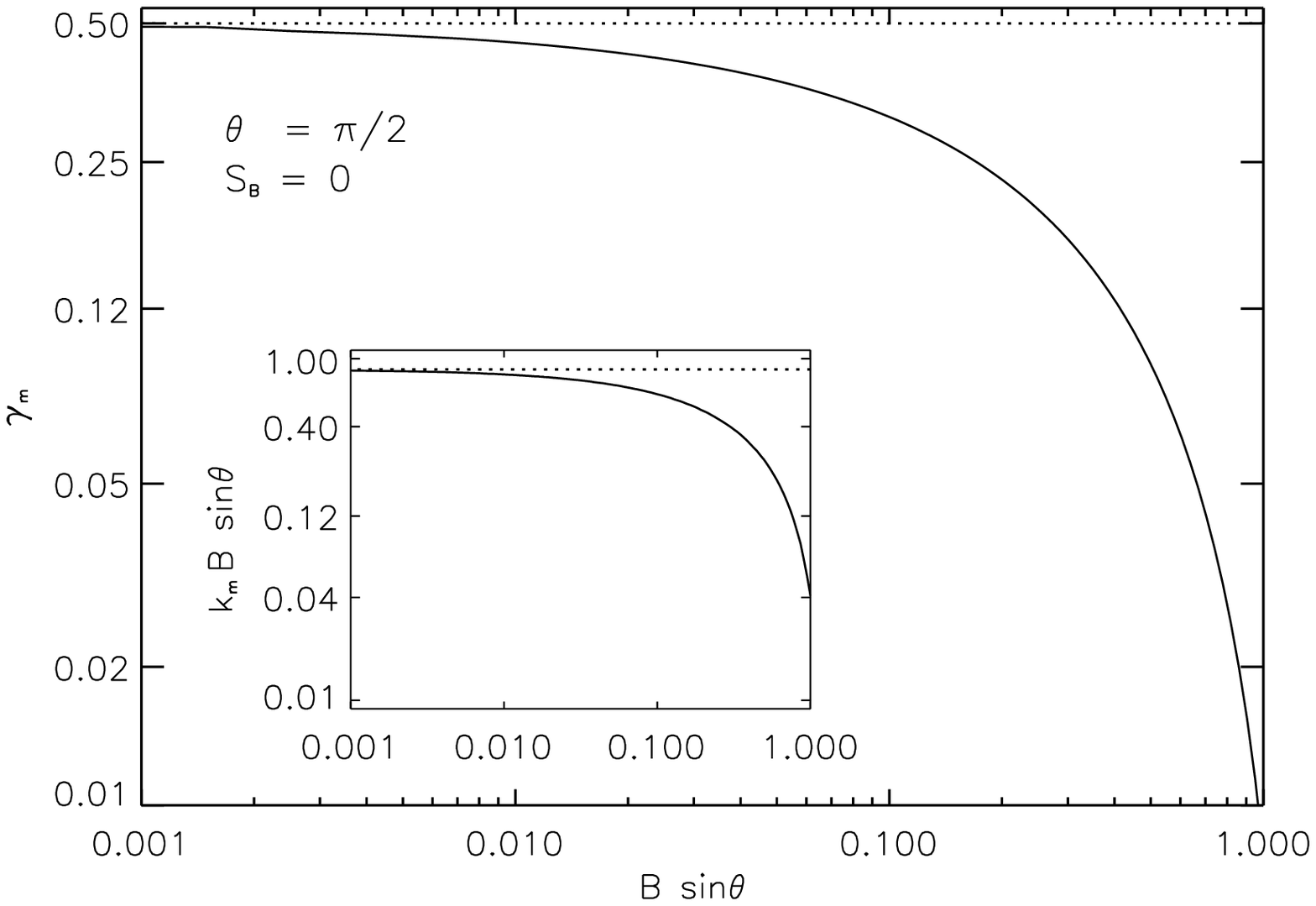}
  \caption{{\it Left panel:} Growth rate of the n=0 (fastest) branch   
    of the inviscid global MRI as a
    function of the vertical Alfv\'en frequency $k \alf \sin \theta $ for  several values of $\alf$
    (time and length are non-dimensionalised with respect to  $r_0$ and  $\om_0$). In the weak field limit, we recover
    the behaviour of the local MRI (dotted line). For all figures, the top-to-bottom order of the varied parameter corresponds to the top-to-bottom order of the main-panel curves. {\it Right panel:}
    Maximum growth rate $\gnm$ as a function of $\alf \sin \theta$. As $\alf
    \rightarrow 0, \, \gnm$ asymptotes to
    the local Oort-A maximum $=1/2$ (dotted line).    Inset: The wavenumber $\km$ at which $\gnm$ occurs as a
    function of $\alf$ along with the local maximum $\sqrt{3/4}$ (dotted line).  Note: In
    the inviscid weak-field limit, two configurations with the same vertical magnetic field will
    exhibit identical behaviour -- 
    Section \ref{sec:tilt_inv}}
  \label{Fig:MRI}
\end{figure*}

Do
these results for our inviscidly stable system, $\mathcal{R}(\gn) =
0$, persist in the presence of the Braginskii viscosity? The answer is
no.

Setting $\theta = 0$ in \eqnt{eq:ODE_0}{eq:v_0}
and making a change of variables, $\dur \rightarrow r^{-1/2}\dur$,
we obtain the simple expression
\begin{equation}\label{eq:energya}
  \gn^2 \odbldif{\dur}{r} - \lt(\gn^2 Q_1 + \gn Q_2 + Q_3\rt)\dur = 0,
\end{equation}
where
\begin{equation}
  Q_1 = \frac{3}{4}\frac{1}{r^2} + \kz^2, \; Q_2 =  \Balbus 
  \frac{B^2 \kz^2 }{r^2}, \; Q_3 = \frac{2 \kz^2}{r^2} (1 + B^2).\nonumber
\end{equation}
We  multiply \eqn{eq:energya} by the complex conjugate of $\dur$, $\dur^\dagger$ and integrate between  $r_0$ and infinity. Boundary terms
vanish, so
 \beq\label{eq:energyb}
\gn^2\lt[\overline{\lt(\odif{\dur}{r}\rt)^2} +
Q_1\overline{|\dur|^2}\rt] + \gn Q_2\overline{|\dur|^2} + Q_3
\overline{|\dur|^2} = 0, \eeq where $\overline{|\dur|^2} =
\int^\infty_{1} \lx{d}r \,\dur \dur^{\dagger} \geq 0$ is non-negative, as is $\overline{(\lx{d}\dur/\lx{d}r)^2}$.

Equation (\ref{eq:energyb}) is a quadratic in $\gn$ whose roots depend
crucially on $\Balbus$. When $\Balbus = 0$ (the inviscid limit), $\gn$
is purely imaginary (neutrally stable, travelling waves), whereas when
$\Balbus> 0$ the result is quite different. In this case
$\mathcal{R}(\gamma)<0 \;\forall\; \Balbus$ and the only question is
whether perturbations are purely damped, or damped and travelling.

It follows that if the system is stable in the absence of the
Braginskii viscosity, it remains so in its presence.


\section{Vertical Field, $\theta = \pi/2$}\label{sec:vertical}


\subsection{Inviscid MRI, $\Balbus = 0$}\label{sec:vertical_inv}

For simplicity, we start by considering the inviscid limit of a purely 
vertical field $\vB = \Bz\vz$, i.e. $\theta = \pi/2, \Balbus = 0$. In this case
\eqnt{eq:ODE_0}{eq:v_0} simplify\footnote{Equations
  (\ref{eq:bessel_root_in}) and (\ref{eq:v_b}) correspond to equations (3.4)
  and (3.5) of \cite{Curry_94} with $a = 1$ and equations (30) and
  (31) of \cite{Dubrulle_93} who have already solved this problem.}:
\beq\label{eq:bessel_root_in}
  \lx{K}_{\li \vn}(k) = 0,
\eeq
with
\beq
  \vn^2 \equiv  -\left[\frac{2 \kz^2}{\gn^2 + \oa^2}\frac{\left(\gn^2 - \oa^2\right)}{\left(\gn^2 + \oa^2\right)} +1 \right].\label{eq:v_b}
\eeq

If $\vn^2<0$, $\lx{K}_{\li \vn} (k)$ has no nontrivial zeros and it follows that for an instability  $\vn^2 >0$
(hence our sign convention in 
\eqn{eq:ODE_0}). This implies $\gn^2$ is bounded
from above by $\oa^2$.

Numerical solutions to \eqna{eq:bessel_root_in} {eq:v_b} are obtained using Newton's method (implemented in Mathematica). There is an unstable solution (and three stable ones) which is shown in Fig. \ref{Fig:MRI}. We find $\gn$ is real, positive, and of order the shear rate. 
The instability is the global MRI, and the $\nn=0, v_0$ branch has the largest growth rate. 

The fastest growing modes have $k \sim 1/\alf$, so in the strong field regime $\alf \lesssim 1, k \gtrsim 1$ the mode is large scale
(physically, smaller scale modes are suppressed by
magnetic tension) and is peaked away from the inner boundary. The result is that $\gnm$ is
reduced below the Oort-A  maximum that occurs
 when  $\alf \sim 1/\kz \ll 1$ and the mode is localised at $r= 1$ \citep{Curry_94}.  

In this weak field regime the problem is amenable to 
asymptotic analysis.  It has been shown by \cite{Cochran_65} and \cite{Ferreira_70,
  Ferreira_08}  that in this  limit, the zeros of $K_{\li\vn}(\kz)$ are given by
\begin{equation}\label{eq:bessel_lim}
  \vn \sim \kz + s_{\nn}  2^{-1}\kz^{\frac{1}{3}}+\ldots,
\end{equation}
where $s_{\nn} = a_{\nn} 2^{\frac{2}{3}}$, $a_{\nn}$ is the modulus of the $\nn^{\lx{th}}$ real negative zero
of the Airy function $Ai$ \citep{Abram} and omitted terms are of the
form $k^b$ with $b < 1/3$. This result can be used to find the
asymptotic-in-$\kz$ form of $\gn$ by inverting equation (\ref{eq:v_b})
to form a bi-quadratic
\begin{equation}
  \gn^4 + 2\left(\oa^2 +\frac{\kz^2 }{1 + \vn^2}\right)\gn^2
  + \oa^2\left(\oa^2 - 2\frac {\kz^2 }{1 + \vn^2}\right)=0,
\end{equation}
 into which we substitute \eqn{eq:bessel_lim}. Retaining the first two terms in $\vn$, we solve exactly for $\gn^2$  
 \begin{equation}
 \gn^2 = - \oa^2 +\lt(1 -  s_{\nn} k^{-\frac{2}{3}} \rt) \lt(-1 \pm \sqrt{\frac{1+ 4 \oa^2}{1 -  s_{\nn} k^{-\frac{2}{3}}  }}\rt),
 \label{eq:vertical}
 \end{equation}
in which the positive root corresponds to the instability. Numerically we find the
fastest growing mode (for a given $\alf, \Balbus$), $\gnm$ and the wavenumber at which 
it occurs $\km$, obey $\partial \gnm/\partial \alf, \partial (\km \alf) /\partial \alf <0$. 

 To draw an analogy with the local
approximation (see Section \ref{sec:local}), $\nn$ that indexes the number of zeros in the domain is
 like a radial wavenumber. For large enough $\nn$, the solutions are plane waves. 
To see this,
we apply a WKB analysis to \eqn{eq:ODE_0} using the small parameter $\kz r/\vn \equiv x/\vn \ll 1$ \citep{Dubrulle_93, Ogilvie_98}. We have
 \beq\label{eq:ODE_WKB}
\odbldif{\dur(x)}{x}+ \frac{\vn^2}{x^2} \dur(x) = 0.  \eeq Demanding
that $\dur$ is real, this has solutions \beq\label{eq:WKB} \dur =
\sqrt{x} \cos (\vn \ln x), \eeq 
and to ensure $\cos(\vn \ln x) = 0$ at
$r = 1$, it must satisfy the boundary condition \beq\label{eq:WKBBC}
\vn \ln k = \lt(\nn + \frac{1}{2}\rt) \pi, \eeq
which determines the spectrum of solutions. 

Combining \eqna{eq:v_b}{eq:WKBBC} we have
\beq
 \gn^2 = - \oa^2 +\alpha \lt(-1 + \sqrt{1 + 4 \oa^2 \alpha^{-1}  }\rt),
 \label{eq:vertical}\eeq
 where $\alpha = \kz^2 (\ln \kz)^2/ \nn^2 \pi^2$ and we have neglected factors of $1/2 \ll \nn$. 

In the large $\nn$ limit, on small enough scales, the mode structure
given by \eqn{eq:WKB} is especially simple.  Reverting to $r$ as our
radial coordinate, we expand \eqn{eq:WKB} about $r = r_1 \sim \order(1)$. To leading
order we find
\beq \label{eq:WKBlocal}\dur = \mathcal{A}\lt(1 + \frac{1}{2}\frac{r}{r_1}
\rt) \cos \lt(\frac{\vn}{r_1} r + \xi\rt),
\eeq
 where $\mathcal{A} = \sqrt{k r_1}$ and $\xi = \vn \ln k r_1$. 
 
 These solutions describes rapidly oscillating modes with frequency $\vn$ and a slowly
varying amplitude $\propto \mathcal{A}$. For sufficiently large $\vn$, 
the solutions are plane waves whose
growth rate decreases with $\nn$. 

\begin{figure*}
  \includegraphics[width=84mm]{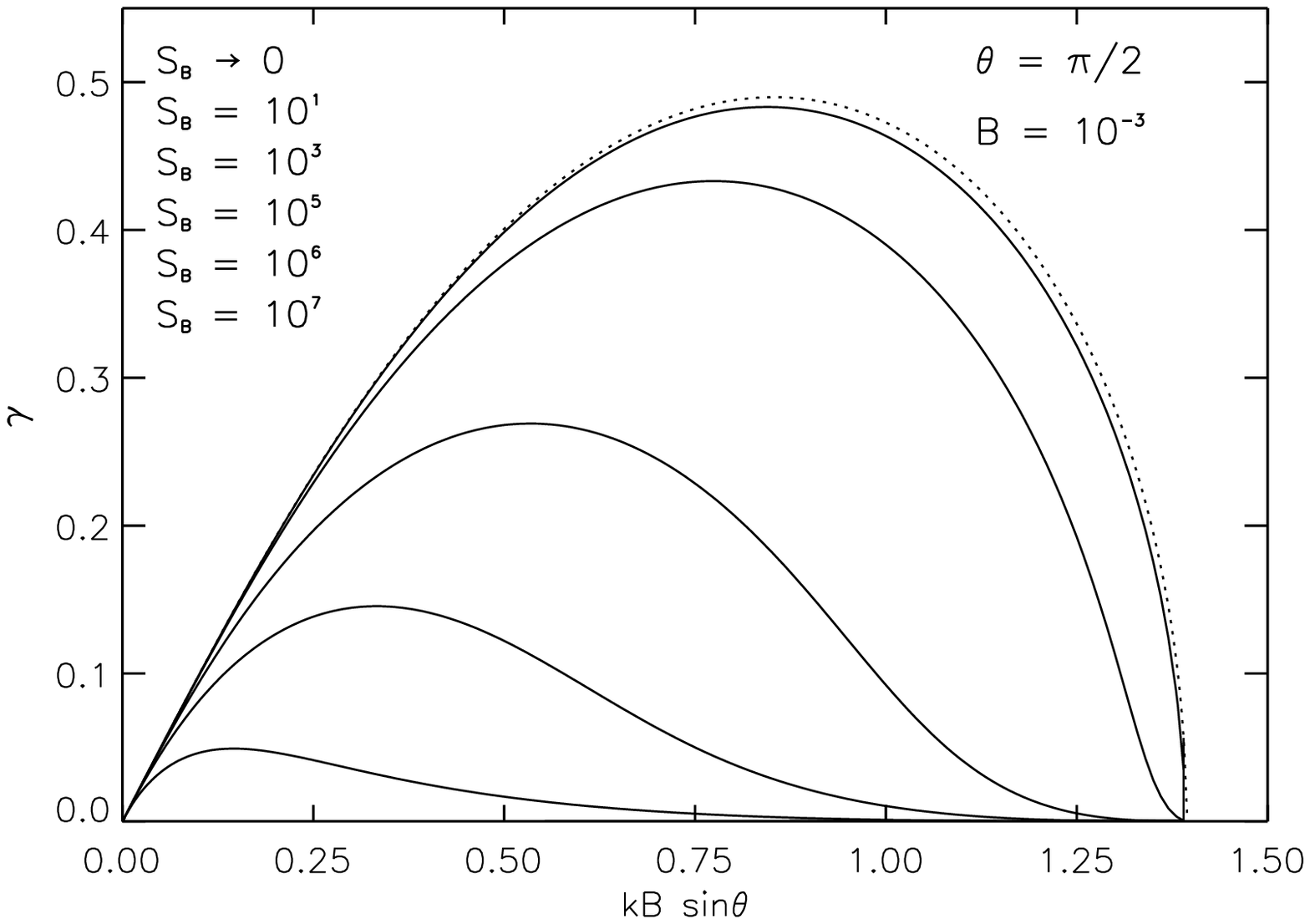}
  \includegraphics[width=84mm]{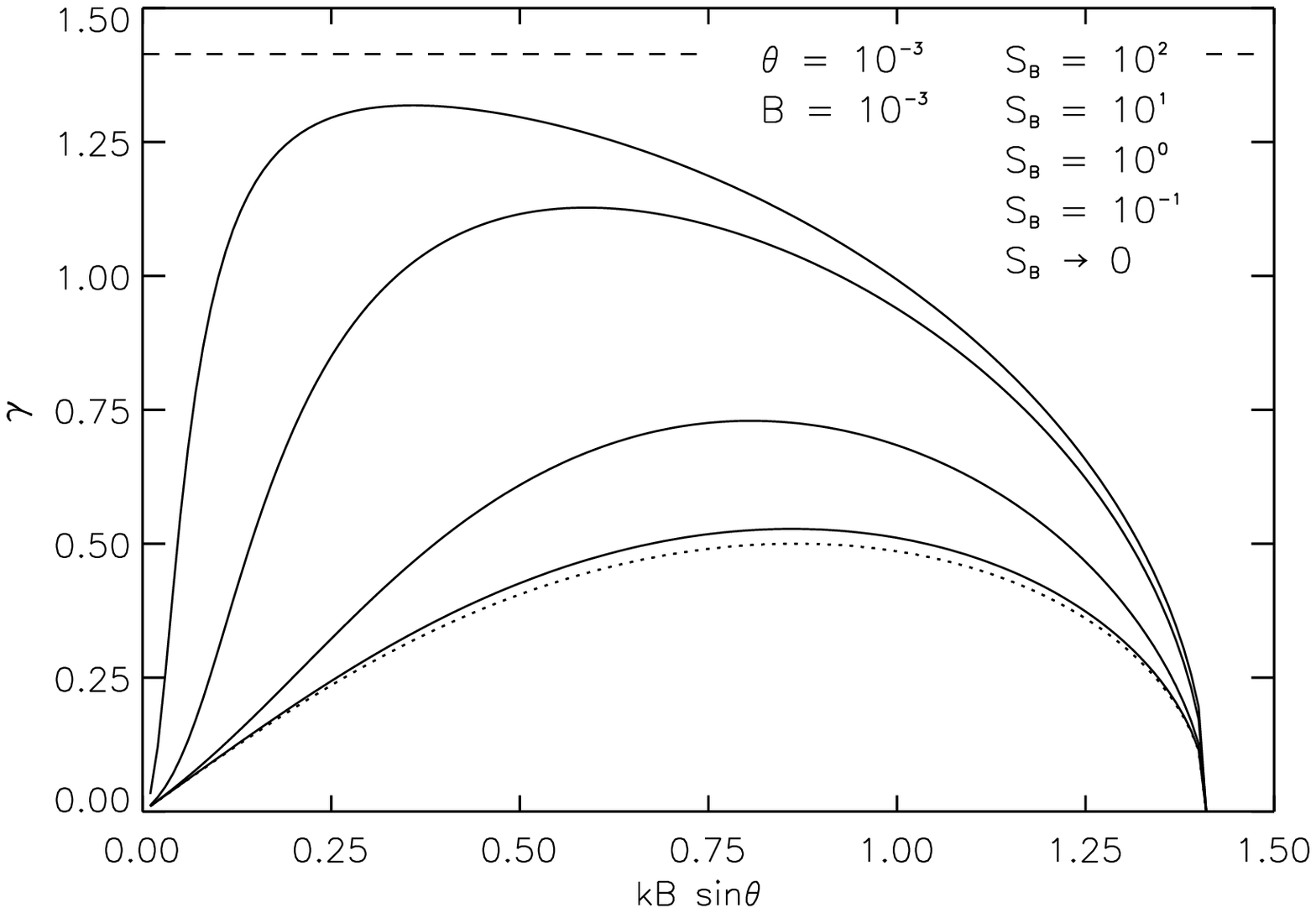}
  \caption{ Growth rate of the n=0 (fastest) branch   
    of the global MRI with Braginskii viscosity for a two different weak, $\alf= 10^{-3}$,
    field configurations and a range of $\Balbus$. {\it Left panel:}
    Vertical Field: For fixed $k$, $\gamma$ decreases as $\Balbus$ increases and is bounded above by the inviscid growth rate
 (dotted line), but the critical wavenumber above which $\gamma < 0$ is independent of $\Balbus$. 
      {\it Right panel:} Slightly pitched field: In
    contrast to $\theta = \pi/2$, for a given $k$, the viscous case $\gamma(\Balbus \neq 0)>\gamma(\Balbus =0)$ inviscid case (dotted), but is less
    than the local Lorentz-force free, or MVI,  limit (dashed). As $\Balbus \to \infty$, $\km \to 0$ but, unlike the vertical case,  $\gnm \to \sqrt{2}.$} 
    \label{Fig:MVI}
\end{figure*}


\subsection{Viscous MRI, $\Balbus \neq 0$}\label{sec:vert_visc}

Allowing for viscosity whilst retaining a vertical field, \eqna{eq:p_0}{eq:v_0} again reduce to a simple form:
\begin{align}
  p^2 &\equiv  \kz^2\frac{ \gn^2 + \oa^2}{\gn^2 + \oa^2 + \Balbus \gn \oa^2 },\label{eq:p_a}\\
  \vn^2 &\equiv  -\left[\frac{2 \kz^2}{\gn^2 + \oa^2 + \Balbus\gn\oa^2 }\frac{\left(\gn^2 - \oa^2\right)}{\left(\gn^2 + \oa^2\right)} +1 \right].\label{eq:v_a}
\end{align}
Unlike the inviscid case, we can
no longer \emph{guarantee} the reality of $p$ as $\gn$ can, and indeed
sometimes does, take complex values. The problem is
generally not of Sturm-Liouville form and so the roots of $\lx{K}_{\li
  \vn}(p)$ are complex. 
  
  Numerically we again find four branches of which three are stable, and one unstable. The
  unstable mode (the only one of interest) is real, so if $\Re (\gn) > 0$ then $\gn \in \Re $. In this case, the problem is Sturm-Liouville. The unstable branch can be traced from the inviscid MRI (and is
  maximized for $n=0, v_0$), Figs. \ref{Fig:Mode}, \ref{Fig:MVI} and \ref{Fig:sing}.
  
  Asymptotic expressions for $\gn$ and $k$ can be found using 
  the  results of \cite{Cochran_65} and \cite{Ferreira_08} for the (complex) roots of \eqn{eq:bessel_root}:
\begin{align}
  \vn &\sim  \pi/( \ln 2 - \gamma_{\rm{Euler}} - \ln p),  &&  p \;\in \;\mathbb{C}, |p| \ll 1, \label{eq:smallroot_limit}\\
  \vn &\sim  p + a_{\nn}  2^{-\frac{1}{3}} p^{\frac{1}{3}}+\ldots, && p \;\in \;\mathbb{C}, |p| \gg 1,\label{eq:root_limit}
\end{align}
where $\gamma_{\rm{Euler}} \simeq 0.58$ is the Euler constant. 

For $\Balbus \gg 1$ we combine \eqna{eq:p_a}{eq:v_a} to get
\begin{multline}
\frac{\gn^4}{k^4}  +  \frac{\gn^2}{k^2}  \lt( \Rem \gn +  \frac{2}{1 + \vn^2}  +  2 \alf^2  \rt)  \\   + \alf^2  \lt(  \Rem \gn - \frac{2}{1 + \vn^2} + \alf^2  \rt) = 0 \label{eq:mestel}
\end{multline}
where $\Rem\!\!= \Balbus \alf^2$ is the inverse of the Reynolds number.  We expand \eqn{eq:mestel} in $\gn^2/k^2 \sim \epsilon \ll 1$ and solve  $\gn = \gn_0 + \epsilon \gn_1 + \ldots$ order by order. To lowest order
we find
\beq
\gn_0 = \Reynolds \lt( \frac{2}{1 + \vn^2} - \alf^2 \rt) \nonumber ,
\eeq 
and to $\order (\epsilon)$ 
\beq 
\gn_1 = -\frac{1 }{\alf^2}  \lt[ \gn_0 + 2 \Reynolds \lt(\frac{1}{1 + \vn^2} +  \alf^2\rt)\rt] \nonumber ,
\eeq
 and so
\beq \label{eq:vert_gamma}
\gn \! \simeq\! \Reynolds \! \lt(\! \frac{2}{1+\vn^2} \!-\! \alf^2 \! \rt) - \! \frac{\gn_0^2}{k^2 \alf^2} \! \lt[\gn_0 + 2 \Reynolds \lt(\! \frac{1}{1 + \vn^2} \! + \! \alf^2 \! \rt) \! \rt] \!\!.
 \eeq

Numerically we find $\km \ll 1$ which, combined with  \eqna{eq:p_a}{eq:smallroot_limit}, implies $\vn\simeq -\pi/\ln k$. It follows that to 
lowest order 
\beq \label{eq:gamma_vert_asym}
\gnm = \Reynolds \, (2 - \alf), \alspace{\Balbus  \gg 1,}
\eeq
and, differentiating \eqn{eq:vert_gamma} with respect to k and neglecting leading order logarithmic variations, 
 \beq \label{eq:k_vert_asym}
 \km \simeq  \frac{2-\alf^2}{\alf \pi} \sqrt{ 2 + \frac{\alf^2}{2} }  \Reynolds \lt( \ln \Rem \rt)^{3/2}, \alspace{ \Balbus \gg  \!1.}
 \eeq

For $\Balbus, \alf \ll 1$, \eqn{eq:root_limit} applies and so $\gn$ is governed by 
\beq 
\lt(\gn^2 + \oa^2   \rt)^2 + \lt[2 -k^{-2}\lt( \Balbus \gn \oa^2 - 2 \rt)\rt]  \lt(\gn^2 - \oa^2   \rt)  =0,\nonumber
\eeq
where we have used $\vn \simeq k \gg 1$. We expand  in $1/k^2  \ll 1$ so $\gn = \gn_0 + k^{-2} \gn_1 + \ldots$, and again solve  order by order. To lowest order
we find $\gamma_0$ satisfies the inviscid equation 
\beq\label{eq:inviscid}
\gamma_0 = \sqrt{-(\oa^2 +1) + \sqrt{1+ 4 \oa^2}}, \alspace{\Balbus \ll 1,}
\eeq
 and at $\order(k^{-2})$
 \beq\label{eq:weak1}
 \gamma_1 = \frac{- \gamma_0^2 \lt( 1 + \Balbus \gamma_0 \oa^2\rt) + \oa^2 \lt( 1-  \Balbus \gamma_0 \oa^2\rt)}{4 (\gamma_0 + \oa)^3 + \gamma_0}, \alspace{\Balbus \ll 1.} 
 \eeq

To see the effect of variations in $n$ on $\gn$ we perform a WKB analysis in which we order $\vn \gg k \gg 1 \gg \Balbus$ so $\ln p/ \vn $ is a
small parameter -- see Section \ref{sec:vertical_inv}. The boundary condition for the viscous modes is then
\beq
\vn \ln \lt[ k \lt( 1- \frac{1}{2} \frac{\Balbus \gn \oa^2}{\gn^2 + \oa^2} \rt)\rt] = \lt(\nn + \frac{1}{2}\rt) \pi. \nonumber
\eeq 
Combining this with \eqn{eq:v_a}, we expand in $\Balbus \ll 1$ so $\gn = \gn_0 + \Balbus \gn_1 + \ldots$. To lowest order $\gamma_0$ is given by the inviscid 
\eqn{eq:vertical}, and to $\order(\Balbus)$
\beq
\gamma_1 = \frac{- \gamma_0 \oa^2 \sqrt{\gamma^2 + \oa^2}}{4 (\gamma_0 + \oa)^3 + 2 \alpha \gamma_0} < 0, \alspace{ \vn \gg 1, \Balbus \ll 1.} 
\eeq  

In the various limits, our asymptotics confirm the numerical results that $\gn \in \Re$, and $\partial \gn/\partial \Balbus < 0$.  Contrasting the results of this section with the inviscid case when $\alf \ll 1$, we have

\begin{align} 	 \label{eq:invlimits}
		&\Balbus \rightarrow 0,  
			&&\gnm \simeq 1/2,   
			&&\km \simeq \alf^{-1} \sqrt{3/4}, \nonumber \\
	&\Balbus \to \infty,  
	&&\gnm \simeq  \Balbus^{-1} \alf^{-2},   
	&&\km \simeq \Balbus^{-1} \alf^{-3} (\ln \Balbus \alf^2)^{3/2}. \nonumber
\end{align}


\section{Pitched Field, $\theta \ll 1$}\label{sec:pitched}


\subsection{Ordering assumptions}

When the magnetic field has both a vertical and an azimuthal
component, the perturbed Braginskii stress tensor
exerts an azimuthal `tension' force on separating plasma elements.
For arbitrary $\theta$, the system is neither
Sturm-Liouville (\eqn{eq:ODE_0} is complex) nor is it amenable to the
kind of polynomial inversion used in Section \ref{sec:vertical}.

However, assuming $\theta \ll 1$ 
matters simplify considerably. Physically, this choice of pitch angle
represents the most realistic non-isolated galactic magnetic field
configuration \citep{Beck_96}.  Mathematically, as we now show, it is
a singular limit that constitutes the stability threshold between
actively damped modes, Section \ref{sec:azimuthal}, and an unstable
configurations that grows even faster than the inviscid global MRI (to
which the following stability threshold also applies)
\beq
  \lx{Damped:}\;\;\theta = 0, \qquad \qquad \qquad \lx{Unstable:}\;\;\theta \sim \eps. \nonumber
\eeq
Here $\eps \ll 1 $ is a small parameter with respect to which we order
the pitch angle and the remaining quantities in \eqn{eq:ODE_0}:
$\gamma, \kz, \lx{d}/\lx{d}r, \alf$ and $\Balbus$.

Letting $\theta \sim \eps$ we can retain only the first few terms in a
series expansion of our trigonometric functions
\beq \cstp{} = 1 -  \frac{\theta^2}{2} + \order(\theta^4),\qquad
\sntp{} =\theta - \frac{\theta^3}{6} +\order(\theta^5).  \nonumber \eeq

To include their effects, we assume a strong magnetic field  $\alf \sim \order(1)$ and
Braginskii viscosity $\Balbus \sim \order(1)$. To retain vertical magnetic tensions $\oa = \kz \alf\sin \theta$
to order 1 (failing to do so removes the high wavenumber cutoff and
leaves the equations ill-posed), we set $\kz \sim 1/\eps$. Then, in
anticipation of unstable modes that grow at the shear rate, we order
$\gamma \sim 1$ too. Balancing terms in \eqn{eq:ODE_0}, we find $\lx{d}/\lx{d}r \sim p \sim k \sim 1/\eps$. 
\begin{figure*}
  \includegraphics[width=84mm]{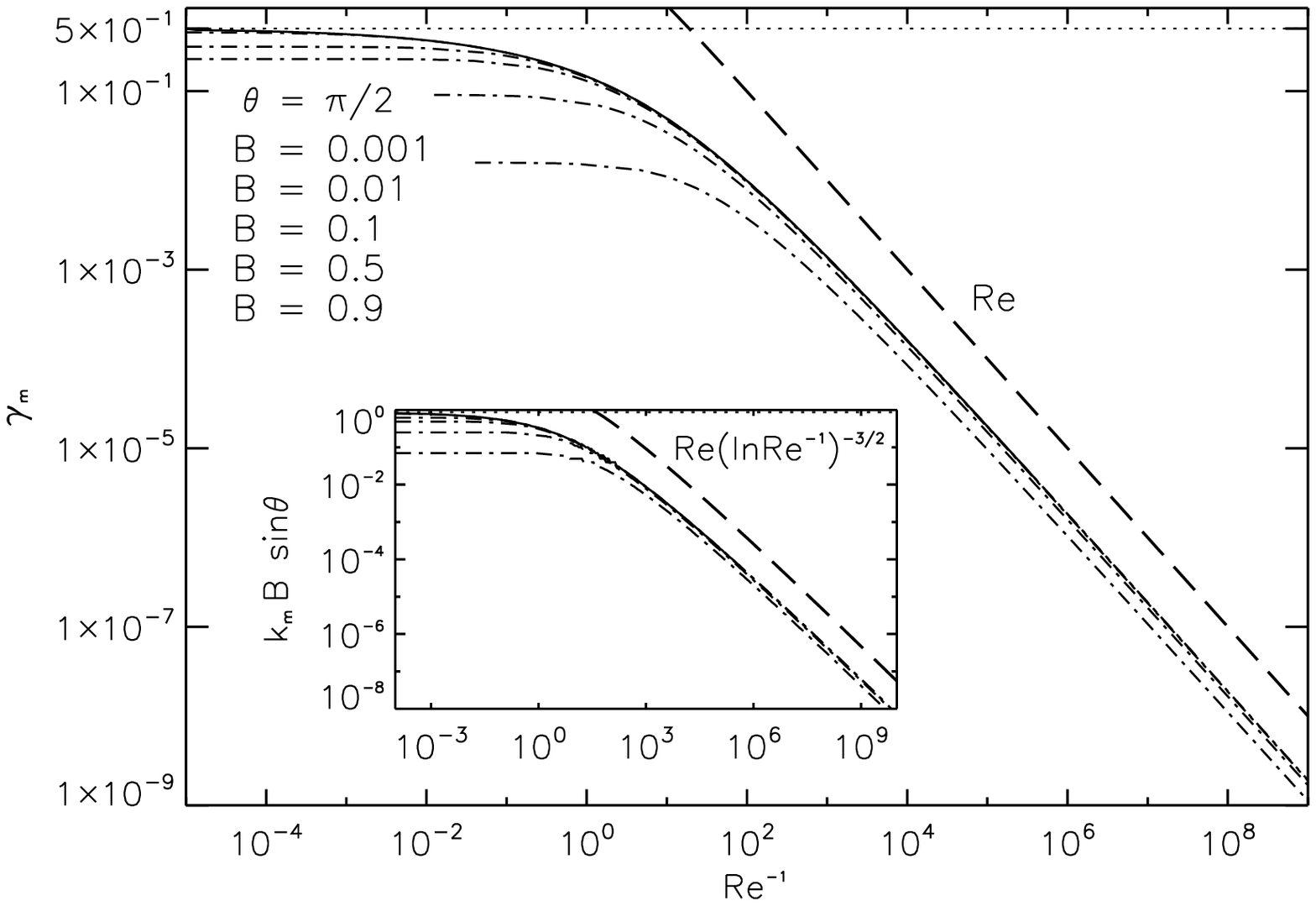}
  \includegraphics[width=84mm]{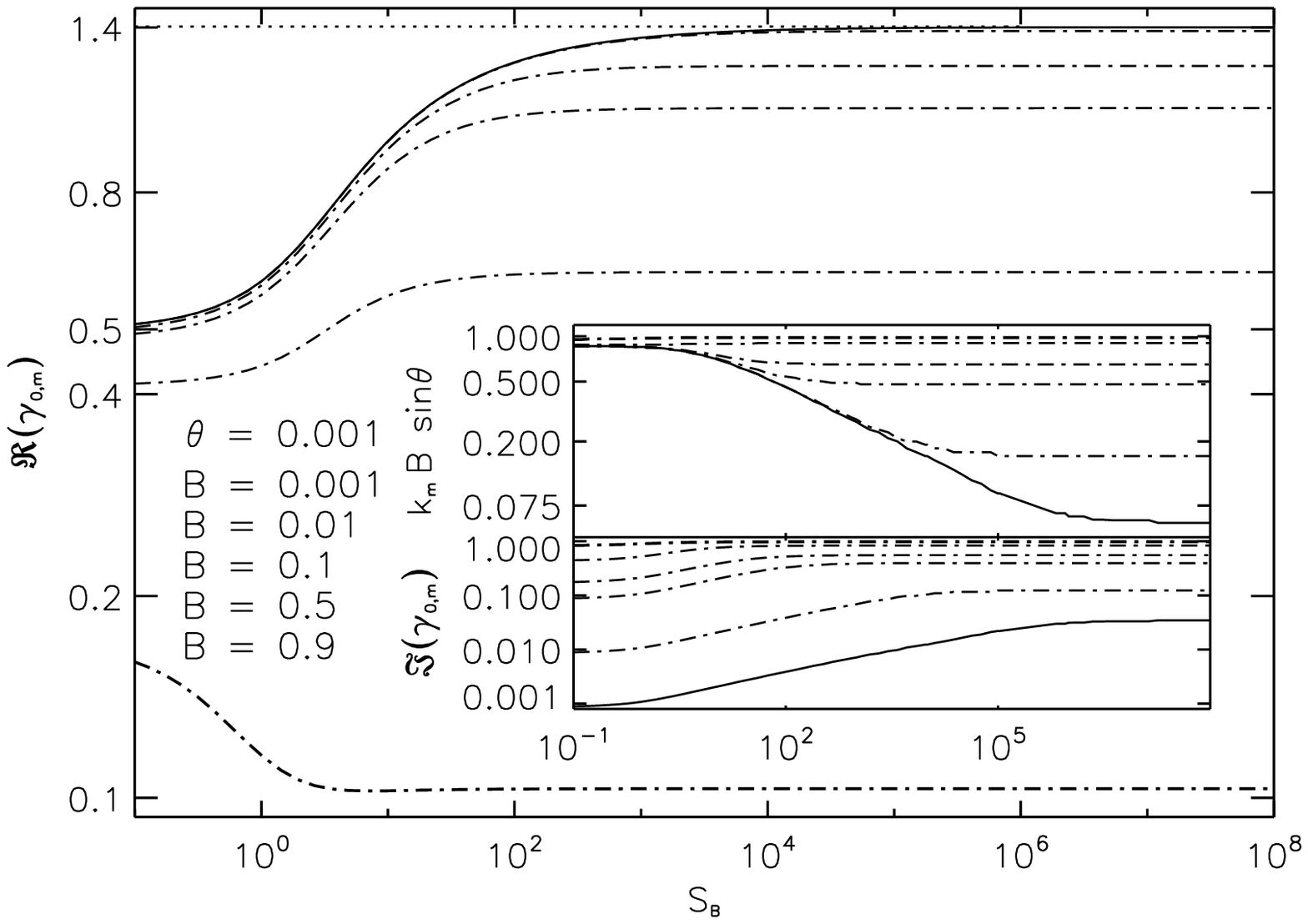}
  \caption{  
  $\Re (\gnm), \Im(\gnm)$ (if it exists), and the corresponding vertical Alfv\'en frequency  $\km \alf \sin \theta$   for two field configurations. {\it Left panel:} Vertical
   field:     $\gnm \in \Re$ and $\km B \sin \theta$  vs. $\Rem \!\! = \Balbus \alf^2$ for a range of $\alf$. These are bounded above by the $\alf \ll1$
   local  limits of $1/2$ and $\sqrt{3/4}$ (dotted lines), and well matched by the asymptotic results (dashed lines) given by \eqna{eq:gamma_vert_asym}{eq:k_vert_asym}. 
     {\it Right Panel}: Slightly pitched,
    $\theta = 10^{-3}$: $\Re(\gnm)$ and $\km B \sin \theta$ are bounded from above by the local $\alf \ll 1$ limit of $\sqrt{2}$ (main panel,
    top dotted line) and the inviscid limit $\sqrt{3/4}$ (top inset,
    dotted line).  The imaginary
    part of the growth rate $\Im(\gnm)$ (bottom inset) is well
    described in the  small and large $\Balbus$ limits  by
    \eqna{eq:intiltb}{eq:gamba}. Note that $\partial \gnm (\alf \sim 1)/\partial \Balbus < 0$ (bottom curve, main panel) demonstrates the combined effects of hoop tension and the Braginskii viscosity.}
\label{Fig:MVImax}
\end{figure*}


In
summary, our orderings are
 \beq \theta \sim \eps, \qquad
\gamma \sim \oa \sim \alf \sim \Balbus \sim 1, \qquad \kz \sim \odif{}{r} \sim
\eps^{-1} \label{eq:ordering}.  \eeq

We apply these scalings to \eqnt{eq:ODE_0}{eq:v_0} and retain the lowest order $\eps$ terms. We find $\dur$ is still governed by
Bessel's equation and, since $p \equiv k$, $\gn$ is determined by 
\eqn{eq:bessel_root_in}  and 
 \beq
 v^2 \equiv -\frac{1}{\ee_1}(\gamma^4 + \rr_3 \gamma^3 + \rr_2 \gamma^2 +
  \rr_1 \gamma + \rr_0 ),\label{eq:v_c}
\eeq
where
\begin{align}
  \ee_1 &= (\oa^2 + \gamma^2)(\oa^2 + \gamma^2+  \Balbus\gamma\oa^2),\nonumber\\
  \rr_3 &=  \Balbus \oa^2\theta^{-2},\nonumber\\
  \rr_2 &= 2 k^2\lt(1 +\alf^2 \rt) - 4 \li \Balbus \oa^2 \kz\theta^{-1},\nonumber\\
  \rr_1 &= - \Balbus\oa^2 k^2 \lt(2 + \alf^2 \rt) - 8 \li \oa \kz^2\alf,\nonumber\\
  \rr_0 &= -2 \oa^2 k^2 \lt(1 + \alf^2 \rt).\nonumber
\end{align}
We start by solving the inviscid problem.


\subsection[]{Pitched inviscid MRI. $\Balbus = 0$}

In the inviscid weak-field regime, we recover  the vertical 
weak-field instability of Section \ref{sec:vertical_inv} and Fig. \ref{Fig:MRI}.  $\alf$ appears via the
vertical Alfv\'en frequency only so $\km(\theta=\pi/2)/\km(\theta=\lx{tilted}) = 1/\theta \gg 1$ so the
unstable mode will be confined to a boundary layer of width $\sim 1/(\alf \theta) \sim \order(\eps^2)$.

If $\alf$ is small but finite (implicitly all orderings are subsidiary to \eqn{eq:ordering}) the governing dispersion relation is the 
complex quartic\footnote{Solutions to this polynomial are considered in detail in \cite{Curry_95} 
and further discussion can be found in \cite{Knobloch_92}.}
\bea
(\gn^2+\oa^2)^2 + 2(1 + \alf^2)(\gn^2-\oa^2) - 8 \li \alf \oa \gn =0.\nonumber
\eea

Writing $\gn$ as a series in $\alf \ll 1$ we find
\beq\label{eq:intilt}
\gn \simeq 2 \li\frac{\alf \oa}{\sqrt{4 \oa^2+ 1}}+  \sqrt{\lt(\sqrt{4 \oa^2 +1}- \oa^2 -1\rt)},
\eeq
with a maximum value 
\beq\label{eq:intiltb}
\gnm = \frac{1}{2} + \li \sqrt{\frac{3}{4}}\alf,
\eeq
which occurs at $\km = \sqrt{3/4} \alf^{-1}$, Fig. \ref{Fig:MVImax}.

\subsection[]{Pitched  MRI with Braginskii viscosity, $\Balbus \neq 0$}
\label{sec:tilt_inv}

\begin{figure}
 \includegraphics[width=84mm]{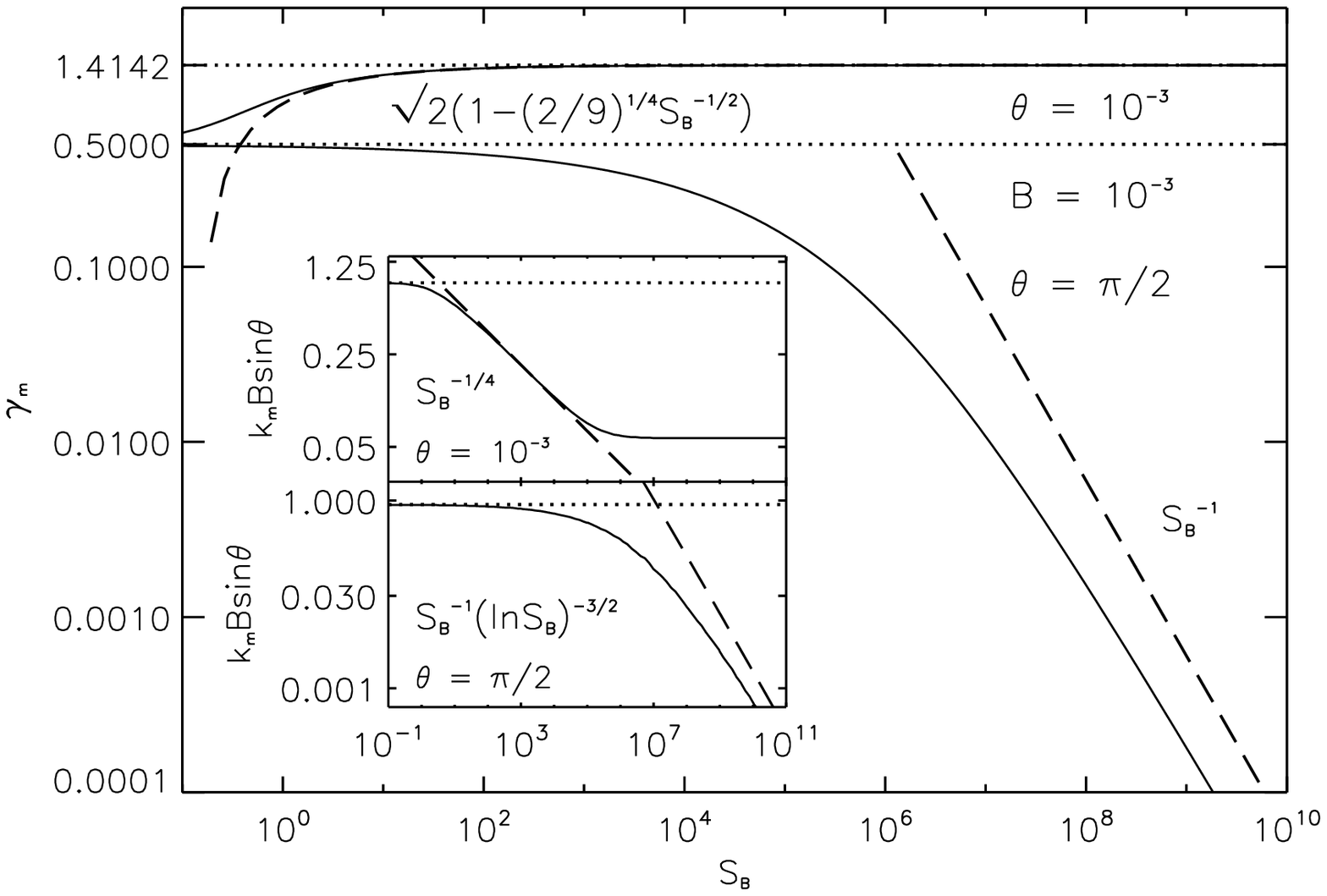}
  \caption{Direct comparison of the $\gnm$ (main panel) and $\km \alf \sin \theta$ (insets) behaviour of the unstable
    mode for $\theta = 10^{-3}$ (top solid line and inset) and $\theta
    = \pi/2$ (bottom solid line and inset) with $\Balbus$ for a weak
    magnetic field, $\alf = 10^{-3}$. The dotted lines are the local maxima, and
    the dashed lines are the asymptotic scalings given by \eqna{eq:gamma_vert_asym}{eq:gmm} for $\gnm$, and
    \eqna{eq:k_vert_asym}{eq:oam} for $\km\alf \sin \theta$. }
\label{Fig:sing}
\end{figure}

In the presence of the Braginskii viscosity we find a complex instability whose real part exceeds that of its inviscid counterpart. Numerical solutions described by 
\eqna{eq:bessel_root_in}{eq:v_c} are shown in Figs. \ref{Fig:MVI}, \ref{Fig:MVImax} and \ref{Fig:sing}. When $\alf \to 0, \Balbus \to \infty$ the growth rate tends to the shear rate $\Re(\gn)\to  \sqrt{2\lx{d}\ln\om/\lx{d}\ln r}=\sqrt{2}$ and, as before, the fastest growing modes occur for $\nn=0$.

Asymptotic expressions for $\gn$ and $k$ can be found by substituting \eqn{eq:bessel_lim} into \eqn{eq:v_c}. For $\alf \ll 1$, $\gn$ is 
governed by 
\beq (\gn^2+\oa^2)^2 + 2(\gn^2-\oa^2)+ \Balbus \oa^2 \gn (\gn^2+\oa^2 -2)=0\nonumber.
\eeq
Differentiating this with respect to $\oa$ 
($k, \alf$ and $\theta$ appear together only  in one
combination and so $\oa$ is treated as a single independent variable), and
setting $\lx{d}\gn/\lx{d}\oa =0$, the stationary points of $ \gn$ obey: 
\beq \Balbus\gn^3 + 2\gn^2 + 2 \Balbus (\oa^2 -1)\gn +
2(\oa^2 -1) = 0.\label{eq:asmax} \eeq
  By considering the
branch  $
\oam$ that maximizes the stationary value  of $\gn$, and introducing $\wm{} =
\oam^2$, we find \beq \wm{} = \frac{\Balbus \gnm ( 2 - \gnm^2) + 2 (1 -
  \gnm)}{2(\gnm \Balbus+ 1)},
\label{as:oa}
\eeq and substituting this into \eqn{eq:asmax} we obtain a
polynomial  whose solutions describe $\gnm$: 
\beq\label{eq:asgov2} 
\gnm^6 - 4 \gnm^4  + 4 (1 - 4\Balbus^{-2})\gnm^2 + 4\Balbus^{-1} (2 \gnm + \Balbus^{-1}-5 \gnm^3)
 = 0.
  \eeq

Because it is a sixth order polynomial, there is no general formulae
for its roots. However, the presence of the asymptotic parameter $\Balbus$ recommends a
series solution. The power of the expansion parameter
 $\Balbus$ depends on whether it is large or small.

Taking $\Balbus\ll 1 $ first, we write 
\begin{align}
\gnm &=\gamma_{m,0} + \Balbus \gamma_{m,1} +  \order(\Balbus^{2}), \label{eq:gseries}\\
\wm{} &= \wmn{0} + \Balbus \wmn{1} + \order(\Balbus^{2}).
\label{eq:oaseries}
\end{align}

Substituting \eqn{eq:gseries} into \eqn{eq:asgov2}, and solving order by order, we find,
to lowest order $\gamma_{m,0} = 1/2$. To next order 
$\gamma_{m,1} = 3/32$ and so
\beq
\gnm = \frac{1}{2} \lt(1 + \Balbus \frac{3}{16}\rt), \alspace{ \Balbus \ll 1.}\label{eq:g_seriesA}
\eeq
Substituting this into \eqna{as:oa}{eq:oaseries}, we find $\oam = \sqrt{3/4} \lt( 1 - \Balbus/48\rt)$, so the most unstable mode is
\beq
 \km =  \frac{1}{\alf\theta} \sqrt{\frac{3}{4}} \lt( 1 - \Balbus \frac{1}{48}\rt) \alspace{ \Balbus \ll 1.} \label{eq:oa_seriesA} 
\eeq

Now taking $\Balbus \gg 1$, we write
\begin{align}
 \gnm &= \gmn{0} + \Balbus^{-1/2}
\gmn{1} + \Balbus^{-1}\gmn{2} + \order(\Balbus^{-3/2}),\label{
  eq:asgam}\\
\wm{} &= \wmn{0} + \Balbus^{-1/2} \wmn{1} + \Balbus^{-1} \wmn{2} +
\order(\Balbus^{-3/2})\label{eq:asoa}.  
\end{align}

Following the same procedure, to lowest  order
  $\gmn{0} = \pm\sqrt{2},0$ and we take the
positive root corresponding to the instability. At first order we
obtain no information, but at
second order we find that $\gmn{1} = \pm
2^{3/4}/3^{1/2}$ .  Taking the negative root (see Fig.  
\ref{Fig:sing}) we have
\begin{equation}
\label{eq:gmm} \gnm =\sqrt{2} \lt( 1 - \lt(\frac{2}{9}\rt)^{1/4}
\!\!\!\! \Balbus^{-1/2}\rt), \alspace{ \Balbus \gg 1.}
\end{equation} 
Substituting this and \eqn{eq:asoa} into \eqn{as:oa}, to lowest order $\wmn{0} =
0$, and at next order 
$\wmn{1} = 2^{5/2}/3^{1/2}$ so $\oam = (2^5/\,9)^{1/8} \Balbus^{-1/4}$,  and 
\begin{equation}\label{eq:oam} 
\km = \frac{1}{\alf\theta}  \lt(\frac{2^5}{9}\rt)^{1/8} \!\!\!\! \Balbus^{-1/4}, \alspace{ \Balbus \gg 1.} 
\end{equation}

Now considering the effect of finite magnetic fields,  $\Re(\gnm)$ (and $\Im(\gnm)$ which is now present) increase with $\Balbus$. Guided by the numerical results,
we assume $|\gn| \sim 1$ and take the leading order balance 
of terms $\propto \Balbus \rightarrow \infty$ in \eqn{eq:v_c}. We find 
\beq
\gn^2 \lt(\oa^2 + \alf^2\rt) - 4 \li \oa \alf \gn + \oa^2(\oa^2 - \alf^2 -2) = 0, \nonumber
\eeq
and the resultant growth rate is
\beq\label{eq:gamba}
\gn = \frac{2 \li \alf \oa \!+ \!\oa \sqrt{\lt(\alf^2 - \oa^2\rt) - 2(\alf - \oa) }}{B^2 + \oa^2},  \alf \lesssim 1, \Balbus \gg 1,
\eeq
which agrees well with the numerics, Fig. \ref{Fig:MVImax}. 

To see the effects of $\nn$ on $\gn$ we adopt the same WKB approach and orderings as Section \ref{sec:vert_visc} where
$\vn$ is now given by the appropriate limit of \eqn{eq:v_c}. We assume $\alf \ll 1$ and expand in $\Balbus \ll 1$ so 
$\gn = \gn_0 + \Balbus \gn_1 + \ldots$.  To lowest order $\gn_0$ is given by \eqn{eq:vertical}, and to $\order(\Balbus)$
\beq
\gamma_1 = \frac{\gamma_0 \oa^2 ( 2 \alpha - \gamma_0^2 - \oa^2)}{4 (\gamma_0 + \oa)^3 + 2 \alpha \gamma_0} >0,
\eeq
so the  Braginskii viscosity  increases the growth rate beyond the inviscid limit.

Contrasting the results of this section with the inviscid case, for $\alf \ll 1$ we have
\begin{align}  \label{eq:tlimit}
			&\Balbus \rightarrow 0,  
			&&\Re(\gnm) \simeq 1/2,   
			&&\km \simeq (\theta \alf)^{-1} \sqrt{3/4}, \nonumber \\
	&\Balbus \rightarrow \infty,  
	&&\Re(\gnm) \simeq \sqrt{2} - \Balbus^{-1/2},   
	&&\km \simeq (\alf \theta \Balbus^{1/4} )^{-1}, \nonumber
 \end{align}
with the caveat that  $\km\to0$ only in the formal limit  $\alf \to 0$ too (see Fig. \ref{Fig:MVImax} for the effect of finite $\alf$ at  large
$\Balbus$).

Considering the effects of hoop tension, we find $\Im (\gn)$ is an order of magnitude greater in the viscous limit. For example, we find
numerically that $\Im(\gn(\Balbus = 10^8)/\gn(\Balbus =0)) \simeq 40$ for $\alf = 10^{-3}$ (which agrees with \eqna{eq:intiltb}{eq:gamba} to 
within half a percent for this and all other examples with $\alf \lesssim 0.5, \Balbus \gg 1$.)


\section{Discussion}\label{sec:discussion}


\subsection{Physical mechanism}\label{sec:mechanism}

We have seen that the orientation of the magnetic field categorically
determines the behaviour of the system. How is this to
be understood physically? 

The most informative explanation comes in the weak field regime where
the role of the magnetic field is twofold. Firstly it facilitates the
generation of pressure anisotropies proportional to its rate of
change, or equivalently (in the collisional regime) $\propto \delta \vbb\vbb:\nabla \vu$.
Because collisions are involved this is a necessarily
dissipative process.  Equations (\ref{eq:mom2}) to (\ref{eq:stress})
yield an energy conservation law
\begin{align}
\odif{}{t} \lt( \frac{\lt< u^2\rt>}{2}+\frac{\lt< B^2\rt>}{2}\rt) =& - 3 \nub \lt<|\vbb\vbb:\nabla \vu|^2\rt>, \nonumber \\
 =& -3\nub \lt< \lt(\odif{ \ln B}{t}\rt)^2\rt>, \nonumber
\end{align}
where $<\cdot>$ are volume averages. 

That is, the Braginskii viscosity damps any
motions that change the magnetic field strength.  This is the first
role of the magnetic field.

The second role of the magnetic field is a geometric one.  Assuming
anisotropies do arise (from changes in the magnetic field strength),
the field's orientation dictates the projection of the anisotropic
stress onto the fluid elements of the plasma, thereby affecting their
dynamics. As we now explain, this fact is crucial in determining the
stability, or lack thereof, and the role global effects have on a
differentially rotating magnetized plasma.

When the field lines have both vertical and azimuthal components (it
is pitched) fluid elements at different heights can exert an azimuthal
stress on each other. The sign of this stress can be either positive
or negative. If the magnetic field is unstable, so its rate of change
is positive, the anisotropic stress acts to oppose any azimuthal
separation of the two elements (this can be identified as the 
fluid version of the stress responsible for the microscopic mirror
force). In this case, like
the MRI, velocity perturbations to fluid elements at different heights
increase (decrease) their angular momentum causing them to move to
larger (smaller) radii. In a system with an outwardly decreasing
angular velocity profile, this leads to an azimuthal separation of the
fluid elements. The associated magnetic field growth ensures the
stress is of the right sign to oppose this separation and this
transfers angular momentum between them in a way that facilitates
further (radial) separation; i.e. an instability (see
\cite{Quataert_02_MRI} for a physical explanation including a spring
analogy). This is the second role of the magnetic field.

In conjunction, the two roles explain our results. In isolated field
geometries where there is no projection of the stress onto fluid
elements at different heights, the Braginskii viscosity does not give
rise to an instability; its only effect is dissipative
\citep{Kulsrud_book,Lyutikov_07, Kunz_10b}. This accounts for the
damping of perturbations in the azimuthal configuration. It also accounts for the
reduced growth rate of the vertical MRI (that depends on the Maxwell,
not the Braginskii, stress), along with the increased radial extent of the mode away from the region of maximum 
shear \citep{Curry_94}. When the field is pitched its dissipative
effects persist, but the free energy contribution from the
differential shear flow will \emph{always} be greater, leading to an
instability (the dissipative effects cannot change the stability
boundary, just the growth rate). Of course the total, equilibrium plus
perturbed, field energy will decrease if viscosity is present (and
viscous heating is not) but, from the perspective of generating
turbulence or transporting angular momentum, this is a secondary
consequence.

Now considering finite magnetic field strengths (and therefore hoop
tension and viscous curvature stresses) we find these modes
become over-stable.  The variation of the travelling 
wave component of the unstable mode is simply a combination of
these two effects, to varying degrees. 


\subsection{Astrophysical example}\label{sec:relevance}

The mathematical results in this paper should be applicable to
any collisional magnetized disc with a decreasing angular velocity profile. However, it is helpful to provide a physical example where our
analysis holds.  Because our theory is developed for a disc with an angular velocity profile $\propto r^{-1}$ we choose the ISM of a spiral galaxy
where a range of analytic and numerical studies already
exist \citep{Kim_03,
  Kitchatinov_04, Dziour_04, Piontek_05, Wang_09}.
  
  Further motivation to study this system comes from
\cite{Sellwood_99} who have argued that in the $\lx{H}\,\lx{I}$ region 
outside the optical disc of a spiral galaxy, the velocity 
dispersion measurements of $\sim 5 - 7~\lx{km~s}^{-1}$ may
be driven by the differential rotation of the disc, mediated by the MRI.  

On this basis, we consider the warm
 ISM in the quiescent regions of a typical
 spiral galaxy where the plasma is generally subject
to a weak magnetic field and an outwardly decreasing angular velocity
profile. It is also magnetized, and so subject to the effects
of the Braginskii viscosity.  This last feature can be seen, and the other relevant 
parameters estimated, by adopting the following set of fiducial
parameters for the ISM \citep{Binney_dynamics, Beck_96, Ferriere_01}. These
are in agreement with $\lx{NGC} 1058$, the well studied face-on disc
galaxy considered by \cite{Sellwood_99}.

Reverting to dimensionalised units for clarity, we have: 

\begin{itemize}

\item Particle number density (ion and electron are the same) \beq\label{eq:density_param}
 n \sim0.3~\lx{cm}^{-3}.  \nonumber \eeq

\item Temperature (we assume ions and electrons are in thermal
  equilibrium) \beq T_i\sim 5\times10^4~\lx{K}; \nonumber \eeq consequently the ion thermal
  speed is \beq
  \vthi=\lt(\frac{2T_i}{m_i}\rt)^{1/2}\sim3\times10^6~\lx{cm~s}^{-1};\nonumber
  \eeq ($T_i$ is in
  erg.)

\item The ion-ion collision frequency (in seconds, assuming $n$ in
  cm$^{-3}$, $T_i$ in K and the Coulomb logarithm $\ln \Lambda = 25$)
  is\footnote{The full expression for the ion-ion collision frequency
    (ion-electron collisions are sub-dominant for a thermalised
    plasma) is given by $\nuii = 4 \sqrt{\pi} n e^4 \ln \Lambda/ 3 m_i^{1/2}
    T^{3/2}$ where $e$ is the elementary charge and $T$ is in erg
    \citep{Brag_65}. } \beq \nuii \sim 1.5 n
  T_i^{-3/2}\sim4\times10^{-8}~\lx{s}^{-1}; \nonumber\eeq  consequently the mean
  free path is \beq
  \mfp=\frac{\vthi}{\nuii}\sim7\times10^{13}~\lx{cm}.
  \label{eq:mfp}\nonumber
  \eeq

\item The typical rotation rate of a spiral galaxy is \beq \om \sim
  5\times10^{-16}~\lx{s}^{-1}.\nonumber\eeq 
  
  A typical value for the outer edge of the optical disc where the turbulence
  cannot be generated by stellar processes is 
   \beq r \sim 3\times10^{22}~\lx{cm},\nonumber \eeq
   and even within the optical disc, at the corotation radius 
   in between the spiral arms, magnetized shear instabilities may 
   	be important (A. Shukurov -- private communication).
	
	 Outside the optical disc 
	a reasonable value for the vertical 
  scale-height is \beq H \sim 10^{21}~\lx{cm}, \nonumber \eeq and
  so the disc is thin. The measured system-scale rotation (not
  turbulent) velocity is \beq U\sim 2\times10^7~\lx{cm~s}^{-1}.\nonumber \eeq

\item The  observed mean magnetic field strengths vary
  between galaxies but on the lower end of the scale are, at the present time \citep{Beck_96} \beq
  B\sim8\times10^{5}~\lx{cm}\,\lx{s}^{-1}\qquad\lx{-
    \;present}. \label{eq:Bpresent} \nonumber \eeq However, if the MRI is the
  dominant turbulence generating mechanism in the ISM, this value
  must represent the saturated state of the magnetic field. Assuming,
  as one must, that present field strengths have been amplified over
  time, at some earlier time they were weaker,
  e.g. \cite{Malyshkin_02, Kitchatinov_04}.

	To ensure the most unstable modes exist at scales $> \mfp$
	so our theory is fluid-like 
  we adopt the following value for the historical `initial' field
  strength and lay aside the problem of where this came from
  \citep{Kulsrud_99, Brandenburg_05},\beq
  B\sim80~\lx{cm} \,\lx{s}^{-1}\qquad\lx{-\;initial}. \nonumber \label{eq:Binitial}
  \eeq
  If we considering a plasma in this era, or the ISM in a galaxy where
  the magnetic field is not saturated, the plasma beta
  is \beq \beta = \frac{\vthi^2}{B^2}\sim 1.3\times10^{9};
  \label{eq:beta}\nonumber
  \eeq the ion cyclotron frequency is \beq \cyc = \sqrt{\frac{4 \pi n}{m_i}}\frac{eB}{c}
  \sim 2\times10^{-6}~\lx{s}^{-1}; \nonumber \eeq ($e$ is the elementary charge,
  $c$ the speed of light) and the ion Larmor radius is \beq \rho_i =
  \frac{\vthi}{\cyc} \sim 1.5 \times 10^{12}~\lx{cm};
  \label{eq:rhoi_ICM}\nonumber
  \eeq and so the plasma is magnetized $\cyc \gg \nuii$ (and will
  become even more so if the magnetic field becomes stronger).

\item The magnetic Prandtl number $P_m$ is huge (and so we
can take the Braginskii viscosity to be the only significant dissipative process)\footnote{The coefficient of
    the Braginskii viscosity is given by $\nub = 0.96 n m_i \vthi^2/\nuii$
    and  $\eta =
    3 \sqrt{2 m_e/\pi} c^2 \ln \Lambda e^2 T^{-3/2}$ where $m_e$ is the
    electron mass and $T$ is in erg \citep{Spitzer_62, Brag_65}.}:
  \beq P_m = \frac{\nub}{\eta} \simeq 7.5\times10^{-6} \frac{T^4}{n} \sim
  7.5\times10^{13}\label{eq:Pm}, \nonumber \eeq where $\eta$ is the coefficient of resistivity and $T$ is in K.

\item The dimensionless parameter $\Balbus = 3 \nub\om/(n m_i \alf^2) \simeq \beta
  \om/\nuii$ is \beq \Balbus \sim 1.1 \times10^{9}~\frac{\om
    T^{5/2}}{B^2} \sim 42,
  \label{eq:Bal}\nonumber \eeq
for the conditions above, so the Braginskii viscosity plays a role. 
\end{itemize}

Our model  of a  
 gravitationally
constrained, differentially rotating disc made up of an isothermal,
incompressible, magnetized plasma fluid is consistent. 
 Neglecting structure in $z$ requires us to restrict our analysis to the galactic plane
by considering vertical scales much less than the  scale height of the disc $H \sim \sound/\om \sim$
where $\sound\sim \vthi$ is the isothermal sound speed. 

Formally, both this and our remaining 
model assumptions can be expressed as a hierarchy of time-scales
which are well satisfied for our set of parameters (we restrict attention to $\gnm$),
\begin{equation}\label{eq:timescales}
  \frac{1}{\cyc} \ll \frac{1}{\nu_{ii}} \ll \frac{1}{\kz \sound}\ll
  \frac{1}{\om}\ll \frac{1}{\om}\frac{\rzero}{H}.\nonumber
\end{equation}
In order of increasing periods of time, these scalings correspond to:
the plasma being magnetized; the plasma being collisional, i.e. a
fluid; the model being uniform in the vertical direction; and the disc
being thin, i.e. rotationally rather than pressure dominated.

The parameters in this section describe a physically realistic regime in which our 
analysis is valid, and potentially important in explaining the gas motions in parts of the ISM. 


\subsection{Relation to the local approximation}\label{sec:local}

Because of its widespread use, we briefly comment on the relation of our results to the local
analysis. In this approximation the coordinate system is Cartesian, the
shear is modelled linearly, and the perturbed quantities are assumed
to vary rapidly in the radial direction (with respect to the
background variations) so they may be written as $ \delta\vc{u}(r)=
\delta\vc{u} \exp[\li (l r + k z) + \gamma t]$, $l r \gg 1$. Under this assumption, a
WKB analysis ignores terms proportional to $1/r$ and replaces
$\lx{d}/\lx{d}r$ by $i l$. The local dispersion relation is \citep{Islam_06, Ferraro_08}:
\begin{equation}
  \label{eq:localdisp}
  \gamma^4 +  \jj_1\gamma^3 + \jj_2 \gamma^2 + \jj_3  \gamma + \jj_4 = 0,
\end{equation}
where
\begin{align}
  \jj_1 = & \Balbus \oa^2\frac{l^2 + k^2\cstp{2}}{l^2 + k^2}, \nonumber\\
  \jj_2 = & 2\lt(\oa^2 +  \frac{k^2}{l^2 + k^2}\rt),\nonumber\\
  \jj_3 = &  \Balbus \oa^2 \left(\oa^2\frac{l^2 + k^2 \cstp{2}}{l^2 + k^2} -2  \cstp{2} \frac{k^2}{l^2 + k^2} \right), \nonumber\\
  \jj_4 = & \oa^2 \left(\oa^2 - 2\frac{k^2}{l^2
      + k^2}\right).\nonumber
\end{align}

In the $k \gg l, l \gg k$ limits for $\theta = 0, \pi/2$ the local and global results are identical (modulo 
different boundary condition-dependent restrictions on the spectrum of allowed modes in the latter case). 

However for
$\theta \ll1, B\lesssim 1$ the analyses differ. Locally we find unstable modes have $\gamma \in \Re$, whilst globally $\gamma$ is complex. Whilst in the inviscid limit $\Im(\gamma) \sim B$ can often be neglected. In the highly viscous limit $\Im(\gamma) \sim \oa B/(\oa^2+ B^2)$ and, as shown in Fig. \ref{Fig:MVImax}, this can be over an order of magnitude greater. This difference may prove important (from
a modelling perspective) for viscous systems which could have been
treated locally, were they inviscid.

Furthermore, inconsistent with the global picture, for $\theta = 0, \pi/2$ and $ l =0$ the local
analysis neglects the Braginskii viscosity. This can be understood from
\begin{align}
 \delta(\vbb\vbb:\nabla \vu)_\lx{G} &= \vphi
\cdot(\vphi\cdot\nabla(\dur\vr)) = \frac{\dur}{r}, \nonumber
&&\theta  = 0,\\
 \delta(\vbb\vbb:\nabla \vu)_\lx{G} &= \vz
\cdot(\vz\cdot\nabla(\duz\vz)) = \li k \duz, 
&&\theta  = \pi/2, \nonumber
\end{align}
whereas locally, assuming  $\nabla \cdot \,\delta \vc{u} = i (k \duz +  l \dur) = 0$ we have
\begin{equation}
 \delta(\vbb\vbb:\nabla \vu)_\lx{L} = 0, \alspace{\theta = 0, \pi/2.}\nonumber
 \end{equation}
 (Here the subscripts G and L stand for global and local.)  Physically, in the global case, a component of the flow is projected
 along the magnetic field direction by the curvilinear geometry (azimuthal case) and the demands of the \emph{global} incompressibility
 condition (vertical case), and so $|\alf|$ changes at linear order, thereby activating the Braginskii viscosity. This is not so in the
 local case.  

One final clarification is worth noting. The local description predicts that the
fastest growing unstable modes ($\theta \neq 0$) occur as $l
\rightarrow 0$. Formally this is inconsistent with the assumption $l
\gg 1/r \neq 0$ that went into deriving \eqn{eq:localdisp}. The local
solutions obtained under the WKB approximation are not self-consistent and should be described by the type
of global solution we have constructed here. However,  having determined the global solutions, we can confirm the $\nn =0$ branch
corresponds to $\gnm$, and so the local analysis is, at least qualitatively, correct in this respect. 


\section[]{Conclusion}\label{sec:Conc}

The nature of the ideal MRI has been well established but
how non-ideal modifications affect it remains an
active topic of research
\citep{Blaes_94, Balbus_01_Hall, Kunz_04, Ferraro_08, Pessah_08,
  Devlan_10}. Of interest to us, and the subject
of this study, has been the global nature of the Braginskii-MHD MRI.

Considering an isothermal, magnetized, collisional disc, perhaps the early ISM,  we have found that for an azimuthal magnetic field with an asymptotically small vertical component, a singular instability
emerges - the magnetized MRI. The growth rate of this instability $\sim \Omega$
 depends on $\Balbus$, a
dimensionless combination of the temperature, shear rate and Alfv\'en speed, and it is up to a factor of
$2 \sqrt{2}$ faster than its unmagnetized counterpart.  In the limit where $\Balbus$ is large, we determined the asymptotic maximum growth rate $ \gnm \simeq \sqrt{2}\lt(1 - 
\Balbus^{-1/2}\rt)$ and corresponding wavenumber $\km \simeq (\theta \alf)^{-1} \Balbus^{-1/4}$.  As the field increases in strength the Larmor radius decreases $\lx{d} \rho_i/\lx{d}t \propto -\Omega$ (so 
the plasma becomes more magnetized) and as the azimuthal component of $B$ increases, purely growing modes mutate into 
over-stabilities whose imaginary part also depends on $\Balbus$. Both of
these factors may be important for the questions posed in Section \ref{sec:intro}, and for the turbulence that will eventually arise. In the
early stages of the instability,  before fully developed turbulence has set it, the instability will generate non-linear pressure anisotropies. What
exact effect these will have is an open question, and so we finish with some thoughts on the subject. 

One major uncertainty in the evolution of magnetized accretion discs is the effect of pressure anisotropy driven micro-instabilities whose growth rates  are generally well in excess of the MRI  e.g. mirror\footnote{In regions where $\lx{d} \ln B/\lx{d}t >0$ slow-wave polarised modes (like the MRI) are unstable
to the mirror with $\gnm \sim \cyc ((p_\perp - p_\parallel)/p)^2$. For the set of parameters listed in Section \ref{sec:relevance} this
is $\simeq 2\cdot 10^{-22} \lx{s}^{-1}$, a factor $10^6$ slower than the MRI. However in many other contexts, e.g. the intracluster medium, the mirror can be up to a factor $\sim10^8$ faster than the macroscopic shear rate 
\citep{hellinger2007comment}.}and firehose \citep{Schek_05}.   Although various models for treating them exist, 
  a complete first-principles theory remains outstanding \citep{Schek_06, Sharma_06, Sharma_07, Rosin_11}. 
  It is crucial for accretion theories to determine the fate of these instabilities fluctuations 
over long (transport) time-scales. This is because they determine the Maxwell and Braginskii stresses that, in turn, dictate the  angular 
momentum transport \citep{Shakura_73}. Even if 
  the pressure anisotropy is pinned at the marginal value for the microscale instabilities in some self-regularizing way, there are 
  further complications that must be addressed.
  
  Specifically, the viscous stress
 generated by the Braginskii viscosity can heat the plasma -- at a
 rate $\propto (p_\perp-p_\parallel)^2$ \citep{Kunz_10b}. Spatial inhomogeneities in the growth rate of the magnetic field, as one would
 expect in a turbulent system, will lead to inhomogeneities in the local pressure anisotropy, both
 in magnitude and sign. In regions of decreasing field strength the firehose would pin the anisotropy at $|(p_\perp - p_\parallel)/p| = 2 /\beta$ and in regions of increasing field strength the mirror would pin it at $|(p_\perp - p_\parallel)/p| = 1/\beta$. The implication of this is that heating in regions of increasing and decreasing magnetic field strength could differ by a factor of $\sim$ four and would occur on
 the decorrelation scale of the turbulent cascade's viscous cut-off --  where the shear is maximized \citep{Schek_06}. If differential heating of this nature does occur then one might expect the non-linear dynamics to be further complicated by 
 magnetized (and unmagnetized) temperature gradient instabilities, and the 
 temperature dependance of both $\nub$ and the micro-instability thresholds  \citep{Balbus_01, Quataert_08, Schek_10}.
  
  Understanding the rich interplay between these  realisations of
magnetized plasma phenomena constitutes an important and, as of yet,
unsolved issue in astrophysics; specifically in accretion discs, but also in galaxies and galaxy clusters. The overall picture is a deeply interconnected
one and the types of processes outlined above are probably pertinent,
to some degree, to most magnetized astrophysical settings. 

To address these issues there is a need for both non-linear simulations that include a wide range of magnetized
physics, and a more complete theory of the transport effects of micro-instabilities. For the first task at least, this work should be useful for 
benchmarking global codes \citep{skinner_10}.


\section*{Acknowledgments}
We thank J.~Binney, T.~Heinemann, M.~Kunz, G.~Lesur, A.~Nahum, G. Ogilvie,  A. Schekochihin, 
A.~Shukurov, J.~Stone and O.~Umurhan for useful discussions and suggestions 
at various stages of this project. 
This work was supported by a STFC studentship (MSR), DOE grant DE-FG02-05ER-25710 (MSR), and  Leverhulme Trust International Network for Magnetized Plasmas (MSR).



\onecolumn

\appendix

\section[]{Linear Stability Analysis}\label{sec:linear}
The linearised \eqnt{eq:mom2}{eq:stress} are, in component form,
\begin{eqnarray}
  \gamma \dur  - 2 \frac{\duphi}{r} &=& - \odif{\delta\Pi}{r} + 
  \li\oa \dbr - 2\frac{\alf \cstp{} \dbphi}{r} - \odif{\dgam}{r} -3 \frac{\cstp{2}}
  {r}\dgam,\label{eq:dur}\\
  \gamma\duphi + \frac{\dur}{r} &=&  \frac{\alf\cstp{}\dbr}{r} +  \li \oa\dbphi
  + 3 \li \kz \frac{\sin 2\theta}{2} \dgam,\label{eq:duphi}\\
  \gamma\duz &=& -\li \kz\delta\Pi + \li\oa \dbz -
  \li \kz (1 - 3\sntp{2})\dgam ,\label{eq:duz}\\
  \gamma \dbr &=& \li \oa \dur,\label{eq:br}\\
  \gamma \dbphi &=& \li \oa \duphi - \frac{\dbr}{r} + \frac{\alf\cstp{}\dur}{r},\label{eq:bphi} \\
  \gamma \dbz  &=&  \li \oa \duz,\label{eq:bz}\\
  \li k \duz  &=& - \frac{1}{r}\odif{(r\dur)}{r},\label{eq:inc}\\
  \dgam &=&\frac{ \Balbus}{3} \alf^2\Bigg[\left(\cstp{2}- \li\frac{\kz}{\gamma}\frac{\sin 2\theta}{2} \right)\frac{\dur}{r} + \li \kz \sntp{2} \duz + \li \kz \frac{\sin 2\theta}{2}  \duphi\Bigg],\label{eq:dgam}
\end{eqnarray}
where  $\gam = \Balbus \alf^2
\vbb\vbb:\nabla \vu$. Equations (\ref{eq:dur})-(\ref{eq:dgam}) form a
closed set which we can combine into a single differential
equation. Eliminating the perturbed magnetic fields, stress, pressure and the
vertical component of the velocity field yields two coupled ordinary
differential equations for $\dur$ and $\duphi$
\begin{eqnarray}
  \cc_0 \odif{\duphi}{r} + \cc_1 \duphi &=& \cc_2 \odbldif{\dur}{r}\label{eq:part1}
  \cc_3 \odif{\dur}{r} +\cc_4 \dur,\\
  \dd_0 \duphi &=& \dd_1 \odif{\dur}{r} + \dd_2 \dur,\label{eq:part2}
\end{eqnarray}
where
\begin{eqnarray}
  \cc_0 &=& \frac{1}{r}\kz^2 2\lt(1 - \li \frac{\oa}{\gamma}\alf\cstp{}\rt)- \li\kz\Balbus\oa^2\cstp{2}\cot\theta,\\
  \cc_1 &=& - \li k \Balbus \oa^2 \frac{\sntp{2}}{2},\\
  \cc_2 &=& -\gamma^{-1}(\gamma^2 + \oa^2) -  \Balbus \oa^2 \sntp{2},\\
  \cc_3 &=& \frac{1}{r}\lt[-\frac{1}{\gamma}(\gamma^2+\oa^2) + \Balbus 
  \oa^2 (\ee_0 - \cot^2\theta)\rt],\\
  \cc_4 &=& \frac{1}{\gamma}(\gamma^2 + \oa^2)\lt(\kz^2 + \frac{1}{r^2}\rt)
  + \frac{1}{r^2} \lt[\frac{2 k^2}{\gamma}\alf \cstp{}\lt(\alf \cstp{} - \li\frac{\oa}{\gamma}\rt) +  \Balbus \oa^2 \ee_0 (\cot^2\theta -1)\rt],\\
  \dd_0 &=& \gamma^2 + \oa^2 +  \gamma \Balbus \oa^2 \cstp{2}, \\
  \dd_1 &=& -\li\kz^{-1}\Balbus \gamma \oa^2 \cstp{}\sntp{},\\
  \dd_2 &=& \frac{1}{r}\lt[\oa\lt( 2\li \alf \cos\theta + \frac{\oa}{\gamma} -
  \frac{\gamma}{\oa}\rt) + \frac{i}{k} \Balbus\gamma \oa^2\cot\theta \ee_0
  \rt],\\
  \ee_0 &=& -\li\kz\gamma^{-1}\sin\theta \cos\theta  + \cstp{2}-\sntp{2}.
\end{eqnarray}

Equations (\ref{eq:part1}) and (\ref{eq:part2}) can then readily be
combined into \eqn{eq:ODE_0}. Solutions to this equation are modified Bessel functions and so the functional forms of the perturbed fields are:
\begin{eqnarray}
  \dur   &=&    \lx{K}_{\li \vn}(p r)  \exp[\li\kz z + \gn t],\label{eq:edur}\\
  \duphi &=& \frac{\dd_1}{\dd_0}\odif{\dur}{r}+ \frac{\dd_2}{\dd_0}\dur,\label{eq:edphi}\\
  \duz &=& \frac{\li}{k} \frac{1}{r} \odif{\left(r \dur\right)}{r}, \label{eq:eduz}\\
  \dbr   &=& \li\frac{\oa}{\gn} \dur, \label{eq:ebr}\\
  \dbphi&=& \li\frac{\oa}{\gamma}\frac{\dd_1}{\dd_0}\odif{\dur}{r} +   \lt[\li\frac{\oa}{\gamma}\frac{\dd_2}{\dd_0}+ \frac{1}{r}\lt(\frac{\alf\cstp{}}{\gamma} - \li \frac{\oa}{\gamma^2}\rt)\rt] \dur,\label{eq:ebphi}\\
  \dbz   &=& -\frac{\oa}{\gn}  \frac{1}{k r} \odif{\left(r \dur\right)}{r},\label{eq:ebz}\\
  \pert{\Pi} &=&  \frac{(\gamma^2 + \oa^2)}{k^2 \gamma} \frac{1}{r} \odif{\left(r \dur\right)}{r} - \Balbus \alf^2 \lt(\frac{1}{3} - \sin^2 \theta \rt)  \lt[\lt(\frac{\dd_2}{\dd_0}-\sntp{2}\rt)\odif{\dur}{r}+\lt(\frac{\ee_0}{r} + \li \kz\frac{\sin 2\theta}{2}\frac{\dd_1}{\dd_0}\rt)\dur\rt].
  \label{eq:eP}
\end{eqnarray}

\end{document}